\documentclass[preprint,review,12pt]{elsarticle}

\usepackage{comment}
\usepackage{url}
\usepackage{booktabs, makecell, tabularx, longtable}
\usepackage[framemethod=default]{mdframed}
\usepackage{xcolor}
\usepackage{longtable}
\usepackage{enumitem}
\usepackage{array}
\newcolumntype{T}{>{\tiny}p}

\journal{Information and Software Technology}

\begin{document}

\begin{frontmatter}

\title{Psycholinguistic Analyses in Software Engineering Text: A Systematic Literature Review}

\author[drexel]{Amirali Sajadi\corref{corresponding_author}}
\affiliation[drexel]{organization={Drexel University},
            city={Philadelphia},
            state={PA},
            country={U.S.A}}
\ead{amirali.sajadi@drexel.edu}
\cortext[corresponding_author]{Corresponding author.}

\author[vcu]{Kostadin Damevski}
\affiliation[vcu]{organization={Virginia Commonwealth University},
            city={Richmond},
            state={VA},
            country={U.S.A}}
\ead{kdamevski@vcu.edu}

\author[drexel]{Preetha Chatterjee}
\ead{preetha.chatterjee@drexel.edu}

    \begin{abstract}
    \textbf{Context:} A deeper understanding of human factors in software engineering (SE) is essential for improving team collaboration, decision-making, and productivity. Communication channels iike code reviews and chats provide insights into developers' psychological and emotional states. While large language models excel at text analysis, they often lack transparency and precision. Psycholinguistic tools like Linguistic Inquiry and Word Count (LIWC) offer clearer, interpretable insights into cognitive and emotional processes exhibited in text. Despite its wide use in SE research, no comprehensive review of LIWC's use has been conducted.
    
    \textbf{Objective:} We examine the importance of psycholinguistic tools, particularly LIWC, and provide a thorough analysis of its current and potential future applications in SE research. 
    
    \textbf{Methods:} 
    We conducted a systematic review of six prominent databases, identifying 43 SE-related papers using LIWC. Our analysis focuses on five research questions: 
    \textit{RQ1. How was LIWC employed in SE studies, and for what  purposes?, RQ2. What datasets were analyzed using LIWC?, RQ3: What Behavioral Software Engineering (BSE) concepts were studied using LIWC? RQ4: How often has LIWC been evaluated in SE research?, RQ5: What concerns were raised about adopting LIWC in SE?}
    
    \textbf{Results:} Our findings reveal a wide range of applications, including analyzing team communication to detect developer emotions and personality, developing ML models to predict deleted Stack Overflow posts, and more recently comparing AI-generated and human-written text.
    LIWC has been primarily used with data from project management platforms (e.g., GitHub) and Q\&A forums (e.g., Stack Overflow). 
    Key BSE concepts include \textit{Communication}, \textit{Organizational Climate}, and \textit{Positive Psychology}.
    26 of 43 papers did not formally evaluate LIWC. 
    Concerns were raised about some limitations, including difficulty handling SE-specific vocabulary.
    
    \textbf{Conclusion:} We highlight the potential of psycholinguistic tools and their limitations, and present new use cases for advancing research on human factors in SE (e.g., bias in human-LLM conversations).
    
\end{abstract}


\end{frontmatter}

\section{Introduction}\label{sec1} 

Understanding human factors in SE is critical to improving collaboration, decision making, and productivity in development teams. The vast amount of textual data generated by stakeholders, developers, and users -- ranging from code reviews to chats -- contains valuable insight into the psychological and emotional states of those involved in software projects. Understanding these factors can reveal key issues in software teams, such as job satisfaction, project participation, and developer attrition~\cite{trinkenreich2022women, raman2020stress, melnik2006comparative}. However, leveraging the full potential of textual data requires the use of specialized tools capable of accurately analyzing unique language patterns and vocabulary used in software engineering~\cite{chatterjee2019exploratory, SE_word_embeddings}.

As one of the most widely used psycholinguistic tools, LIWC has played a significant role in advancing research in many scientific fields \cite{lyu2023detecting, bell2006variations, neysari2016monitoring, simmons2005pronouns, ziemer2017using, gulliver2021assessing, oc2023luxury}. In software engineering studies, LIWC has been adopted and used repeatedly by researchers since 2007 until now, providing insight into developer collaboration, the emotional tone of project discussions, the overall dynamics of software teams, etc. While previous studies have examined the use of opinion mining and sentiment analysis tools in SE~\cite{lin2022opinion}, this study focuses on psycholinguistic tools due to their ability to target and interpret complex psychological constructs. In particular, we explore the significant role of LIWC in the analysis of SE-specific language toward gaining insights into various psychological and social factors affecting the daily tasks of software engineers. Through a systematic paper selection process, we identified 43 papers that study SE-related topics with the help of LIWC. We closely examined each of these papers to answer the following research questions:
\begin{itemize}[leftmargin=*]
    \item \textbf{RQ1: How was LIWC employed in SE studies, and for what specific activities and purposes?}
    \\ We found that LIWC has been predominantly used to study \textit{Team Management} in SE, and applied in tasks such as personality detection, emotional tone analysis, and quality assessment of Stack Overflow posts. Moreover, LIWC has been used both directly and indirectly in these tasks, with the latter including the use of LIWC's output in feature sets of machine learning models, enabling tasks such as predicting the likelihood of a post’s deletion on Stack Overflow.
    
    \item \textbf{RQ2: What textual data/datasets were analyzed by LIWC?}
    \\ We observed a wide range of textual data types and datasets analyzed using LIWC, such as Q\&A posts on  Stack Overflow and Kaggle, developer communication on IRC chat channels, and email archives. This diverse set of data sources highlights LIWC’s flexibility in processing different forms of communication in the SE domain.
    \item \textbf{RQ3: What Behavioral Software Engineering (BSE) concepts were studied using LIWC?}
    \\ The core BSE concepts studied using LIWC include team dynamics, collaboration, motivation, and developer productivity. Understanding these concepts is critical, as they offer insights into how psycholinguistic tools like LIWC contribute to solving practical SE challenges, particularly by analyzing how communication and collaboration impact team performance.
    \item \textbf{RQ4: How often and in what way(s) has LIWC been evaluated in a software engineering context?}
    \\ We identified 17 studies that conducted formal evaluations of LIWC in SE contexts. In our review, evaluation refers to any effort by researchers to assess LIWC’s effectiveness or reliability within their studies. These evaluations varied in scope but provided validation of LIWC's role in SE research.

    We identified 17 studies that conducted evaluations of LIWC in SE contexts. In our review, evaluation refers to any effort by SE researchers to assess LIWC’s effectiveness or reliability. Methods included comparisons with ground-truth data and other forms of methodological validation. These evaluations varied in scope but provided evidence supporting LIWC’s use in SE research.
    
    \item \textbf{RQ5: What concerns or limitations were mentioned by the researchers adopting LIWC in the software engineering domain?}
    \\ We categorized concerns raised about LIWC into two main areas: (1) inherent limitations of LIWC as a lexicon-based tool, e.g., limited language support, and (2) challenges with its adoption in SE, particularly in handling domain-specific vocabulary. These limitations suggest that while LIWC is valuable, there is room for improving its SE-specific vocabulary. 
\end{itemize}

By outlining LIWC's role in software engineering research, we provide researchers with the insights needed to apply LIWC effectively. This study not only highlights its current applications and limitations, but also encourages the community to enhance and integrate LIWC with other tools for a more holistic understanding of software engineering practices. We discuss opportunities for improving LIWC's use and propose new approaches, including the integration of LIWC with LLMs, to advance its application in software engineering research.

\section{Background}

\subsection{Psycholinguistics}
Psycholinguistics is an interdisciplinary field with roots in psychology and linguistics, investigating the complex connection between the mind and language. More specifically, this discipline seeks to uncover the workings of the human brain when individuals engage in language, be it written or spoken \cite{edition2008psychology, harley2013psychology, levelt2013history}.

Psycholinguistics explores a wide range of topics and intersects with several scientific domains. It ventures into the field of neuroscience, studying the physiology of the human brain to understand the functions behind its ability to acquire and use language effectively. Psycholinguists also study the psychological facets of language. Through experimentation, they discover the processes that link language to human cognition, for instance language's complex interplay with short-term and long-term memory \cite{traxler2011handbook}.

In addition, the history of psycholinguistics has deep philosophical roots that have sparked transformative moments in the evolution of science. For instance, the development of psycholinguistics contributed to research on language acquisition, which itself has broadened our understanding of the nature of human learning \cite{fletcher1995handbook}. Similarly, psycholinguistics influenced the shift in the field of psychology from the study of behaviorism to the study of mental processes, for example, language comprehension and generation \cite{miller2003cognitive}. A pivotal moment was Noam Chomsky's critique of behaviorism. The movement he initiated, known as the Chomskyan revolution, significantly affected the field, leaving a mark on the landscape of psycholinguistics \cite{harris1994chomskyan}.

\subsection{Text Analysis}
According to Neuendorf \& Kimberly, \textit{content analysis} can be succinctly defined as ``the systematic, objective, quantitative analysis of message characteristics." \cite{neuendorf2017content}. As a specific type of content analysis that focuses on text-based data, text analysis finds application in various domains through tasks including sentiment analysis, topic modeling, information extraction, text summarization, and language translation.

Today, a plethora of text analysis algorithms and software applications are readily accessible. Text analysis algorithms, however, predate the recent breakthroughs in machine learning, deep learning, and natural language processing techniques that have enabled the recent advancements in the quality of text processing. A distinctive characteristic of the prior generation of statistical text analysis algorithms is their utilization of dictionaries, often multiple, created and used for the analysis of text. Dictionaries provide a mapping between specific words in the language and the relevant concepts under study. For instance, the Linguistic Inquiry and Word Count (LIWC) tool leverages custom dictionaries for the measure of the expression of a specific psychological concept in the text, i.e., as a psycholinguistic analysis tool.

\subsection{Linguistic Inquiry and Word Count (LIWC)}
James W. Pennebaker et al. initially developed Linguistic Inquiry and Word Count (LIWC) \footnote{https://www.liwc.app} to automate parts of the text analysis process they intended to use for analyzing essays written by people who have experienced trauma. Since then, LIWC has been used across different disciplines to analyze text and, among other use-cases, extract psychological information from the text \cite{tausczik2010psychological}.

In a top-down fashion, LIWC's dictionaries categorize words into predetermined categories, mapping them to important psychological constructs and theories.
 The categories cover a wide range of linguistic and psychological dimensions, including cognition, affect, social processes, culture, and motives, to name a few. Table \ref{tab:all_categories} lists all the categories directly mentioned in our selected studies that are still available in the 2022 version of the LIWC. 
 
To improve clarity, we have divided LIWC categories in Table \ref{tab:all_categories} into three subsections as follows: (1) Summary Variables, (2) Text Statistics, and (3) Lexical Variables. The values for \textit{Text Statistics} and \textit{Lexical Variables} are calculated based on the percentage of the words in the text that fall under them. However, \textit{Summary variables} are derived from the percentage of words in other LIWC categories. The Summary Variables stand as the only non-transparent dimensions within the LIWC program, as the precise calculations leading to the values assigned to these categories remain unknown. In Table \ref{tab:all_categories}, we include the explanations or common words listed in each category. For example, under \textit{Lexical Variables} section, ``Drives'' category includes words that are intended to capture motivational language related to affiliation, achievement, and cognitive engagement (e.g., \textit{we}, \textit{know}, \textit{work}).

Based on the word count, i.e., the frequency of specific words in a body of text, and the percentages of words falling into each of the categories, LIWC provides insight into psychological and emotional content in text. Over the years, various studies have validated the correlation between LIWC’s output and a wide range of psychological concepts. In a 2010 publication, Tausczik \& Pennebaker pointed out these correlations by referencing studies that used tools such as surveys, personality assessments, and self-reports to analyze participants' linguistic patterns \cite{tausczik2010psychological}.

\AtBeginEnvironment{longtable}{\scriptsize} 
\begin{longtable}{p{0.45\textwidth} p{0.45\textwidth}}
\caption{LIWC-22 Categories Mentioned in Our Study Subjects.}
\label{tab:all_categories} \\
\hline
\textbf{Category} & \textbf{Description / Frequent Exemplars} \\
\hline
\endfirsthead

\hline
\textbf{Category} & \textbf{Description / Frequent Exemplars} \\
\hline
\endhead

\hline
\endfoot

\hline
\endlastfoot
\\
\multicolumn{2}{c}{\textbf{Summary Variables}} \\ \\
Analytical thinking & Measure of formal/logical thinking \\
Clout & Measure of language indicating confidence and status \\
Authentic & Measure of perceived honesty and genuineness \\
Emotional tone & Measure of of positive vs. negative affect \\ \\

\multicolumn{2}{c}{\textbf{Text Statistics}} \\ \\
Word count & Total number of words \\
Words per sentence & Average words per sentence \\
Big words & Percentage of words 7 letters or longer \\
Dictionary words & Percentage of words matched in LIWC dictionary \\
Total function words & Total number of \textit{function words} ({e.g.}, \textit{the}, \textit{to}, \textit{and}, \textit{I}) \\ \\

\multicolumn{2}{c}{\textbf{Lexical Variables}} \\ \\

1st person singular & \textit{I}, \textit{me}, \textit{my}, \textit{myself} \\
1st person plural & \textit{we}, \textit{our}, \textit{us}, \textit{let’s} \\
2nd person & \textit{you}, \textit{your}, \textit{u}, \textit{yourself} \\
3rd person singular & \textit{he}, \textit{she}, \textit{her}, \textit{his} \\
3rd person plural & \textit{they}, \textit{their}, \textit{them}, \textit{themsel*} \\
Articles & \textit{a}, \textit{an}, \textit{the}, \textit{alot} \\
Prepositions & \textit{to}, \textit{of}, \textit{in}, \textit{for} \\
Adverbs & \textit{so}, \textit{just}, \textit{about}, \textit{there} \\
Conjunctions & \textit{and}, \textit{but}, \textit{so}, \textit{as} \\
Auxiliary verbs & \textit{is}, \textit{was}, \textit{be}, \textit{have} \\
Common verbs & \textit{is}, \textit{was}, \textit{be}, \textit{have} \\
Impersonal pronouns & \textit{that}, \textit{it}, \textit{this}, \textit{what} \\
Numbers & \textit{one}, \textit{two}, \textit{first}, \textit{once} \\
Negations & \textit{not}, \textit{no}, \textit{never}, \textit{nothing} \\
Positive emotion & \textit{good}, \textit{love}, \textit{happy}, \textit{hope} \\
Negative emotion & \textit{bad}, \textit{hate}, \textit{hurt}, \textit{tired} \\
Anxiety & \textit{worry}, \textit{fear}, \textit{afraid}, \textit{nervous} \\
Anger & \textit{hate}, \textit{mad}, \textit{angry}, \textit{frustr*} \\
Sadness & \textit{: (}, \textit{sad}, \textit{disappoint*}, \textit{cry} \\
Social processes & \textit{you}, \textit{we}, \textit{he}, \textit{she} \\
Work & \textit{work}, \textit{school}, \textit{working}, \textit{class} \\
Achievement & \textit{work}, \textit{better}, \textit{best}, \textit{working} \\
Leisure & \textit{game*}, \textit{fun}, \textit{play}, \textit{party*} \\
Money & \textit{business*}, \textit{pay*}, \textit{price*}, \textit{market*} \\
Time & \textit{when}, \textit{now}, \textit{then}, \textit{day} \\
Past focus & \textit{was}, \textit{had}, \textit{were}, \textit{been} \\
Present focus & \textit{is}, \textit{are}, \textit{I’m}, \textit{can} \\
Future focus & \textit{will}, \textit{going to}, \textit{have to}, \textit{may} \\
Visual & \textit{see}, \textit{look}, \textit{eye*}, \textit{saw} \\
Auditory & \textit{sound*}, \textit{heard}, \textit{hear}, \textit{music} \\
Fillers & \textit{rr*}, \textit{wow}, \textit{sooo*}, \textit{youknow} \\
Insight & \textit{know}, \textit{how}, \textit{think}, \textit{feel} \\
Drives & \textit{we}, \textit{our}, \textit{work}, \textit{us} \\
Discrepancy & \textit{would}, \textit{can}, \textit{want}, \textit{could} \\
Tentative & \textit{if}, \textit{or}, \textit{any}, \textit{something} \\
Certitude & \textit{really}, \textit{actually}, \textit{of course}, \textit{real} \\
All-or-none & \textit{all}, \textit{no}, \textit{never}, \textit{always} \\
\end{longtable}

For instance, they highlight how the use of words under the \textit{1st person singular} category e.g., ``I", ``me", and ``mine" reflects self-focus. Psychological concepts like self-reflection, insecurity, anxiety, and even depression have been shown to correlate with increased use of \textit{1st person singular} words in various contexts, as individuals experiencing these states tend to center their attention inwardly, which is reflected in their language use. These findings were established through empirical research that connected participants' word usage with their self-reported psychological states, providing strong evidence of the links between language patterns and psychological phenomena.

Since the initial release of LIWC, multiple versions of it have been created, each characterized by a larger dictionary and a more extensive set of interpretable dimensions. LIWC2015, for instance, uses dictionaries with as many as 4500 words \cite{pennebaker2015development}, while LIWC-22's internal dictionary contains 12000 words \cite{boyd2022development}. Notably, new  functionalities were added to the 2022 version of LIWC that have been absent in the previous versions. For instance, LIWC-22 offers \textit{narrative arc}, \textit{word frequency and word clouds}, and a \textit{dictionary workbench} designed to streamline the cumbersome process behind the creation of custom dictionaries \cite{boyd2022development}. Lists of the dimensions and capabilities included in each version of LIWC can be found in the corresponding published journal articles or manuals, e.g., \url{https://www.liwc.app/help/psychometrics-manuals}.

\section{Related Work}

\subsection{Systematic Literature Reviews of Behavioral Analyses in Software Engineering}

Several SLRs were conducted to identify the use of behavioral analyses in SE. The scope of studies incorporating behavioral analysis in SE is extensive, covering topics such as sentiment and emotion detection \cite{lin2022opinion, 9402078, sajadi2023interpersonal, sajadi2023towards, imran2024uncovering}, personality assessment \cite{felipe2023psychometric, yarkoni2010personality, 10.1145/3611643.3613077}, and job satisfaction \cite{melnik2006comparative, pardee1990motivation}. Researchers employ a wide range of methodologies, from traditional tools like questionnaires and interviews to automated Machine Learning (ML) models, to answer their research questions. This diversity highlights the vast potential for further exploration to deepen our understanding of the human element in the software engineering process.

Lenberg et al.'s comprehensive literature review on behavioral software engineering, for example, demonstrates the variety of approaches available for studying human factors in SE tasks~\cite{lenberg2015behavioral}. Their review points to several key research areas—such as work and organizational psychology, the psychology of programming, and behavioral economics—that contribute to the broader field of behavioral software engineering. This interdisciplinary approach highlights the significance of human behaviors and their influence on software development \cite{lenberg2015behavioral}.

Numerous studies have also explored the impact of various human factors within development environments, along with methods for their detection. Lin et al. conducted a systematic literature review (SLR) focusing on opinion mining in software development, including tasks such as sentiment analysis, emotion detection, and politeness detection \cite{lin2022opinion}. Another survey study looked at empirical research on developer emotions between 2005 and 2018 \cite{sanchez2019taking}. Felipe et al. conducted a similar review, examining the psychometric tools used to infer personality types in software engineering research \cite{felipe2023psychometric}. Studies of this nature frequently identify LIWC as a tool for detecting the psychological concepts and human factors they examine. To date, however, no prior work has systematically reviewed how LIWC itself has been adopted, evaluated, and applied across the entire body of software engineering research. While other SLRs focus on broader psychological or behavioral techniques, our study is the first to concentrate exclusively on the role and methodological implications of LIWC within software engineering.
 
\subsection{Use of LIWC Outside Software Engineering }
Through a variety of applications, LIWC has demonstrated a unique ability to discover insights about individuals through their written or spoken words. LIWC remains a popular tool in diverse domains of study despite certain limitations inherent to a dictionary-based approach, such as its inability to account for sarcasm or negation in text. LIWC remains widely used across research domains, with 810 citations in 2023 alone, highlighting its continued relevance and broad impact in various fields. 

The work of Pennebaker et al. reflects LIWC's versatile use-cases~\cite{pennebaker2001linguistic}. For instance, LIWC has been frequently used to assess  the emotional content of text, performing tasks such as sentiment analysis and the detection of specific emotions \cite{kahn2007measuring}. Additionally, it has been used to assess the personality traits of individuals who contribute to the text \cite{felipe2023psychometric}. In psychology research, many have used LIWC to detect depression \cite{lyu2023detecting, lumontod2020seeing} and to study how the language use of couples relates to their marital satisfaction and the overall health of the marriage \cite{neysari2016monitoring, simmons2005pronouns}. LIWC has also been used to analyze text with political content to assess the public's opinion on political topics and to find common linguistic patterns in speech of famous politicians \cite{kangas2014can, tumasjan2010predicting, apriyanto2020personality}. Other researchers have analyzed psychological concepts within a wide range of texts, including suicidal poems~\cite{stirman2001word}, Al-Qaeda transcripts \cite{pennebaker2008computerized}, and social media posts written by users dealing with the impacts of the Covid-19 pandemic \cite{ashokkumar2021social}. The variety and quantity of these studies have showcased LIWC's unparalleled ability to analyze human emotions, thoughts, and intentions encoded within language.

Despite LIWC’s wide adoption across many disciplines, and its frequent appearance in SE studies, no existing review has examined its specific applications, limitations, and potential, specifically within the context of software engineering. This gap motivates the need for a focused SLR that examines LIWC’s contributions to SE research, an effort we undertake in this study.

\section{Review Methodology}
\label{methodology}

\begin{figure}[tb]
  \centering
  \includegraphics[width=.96\linewidth]{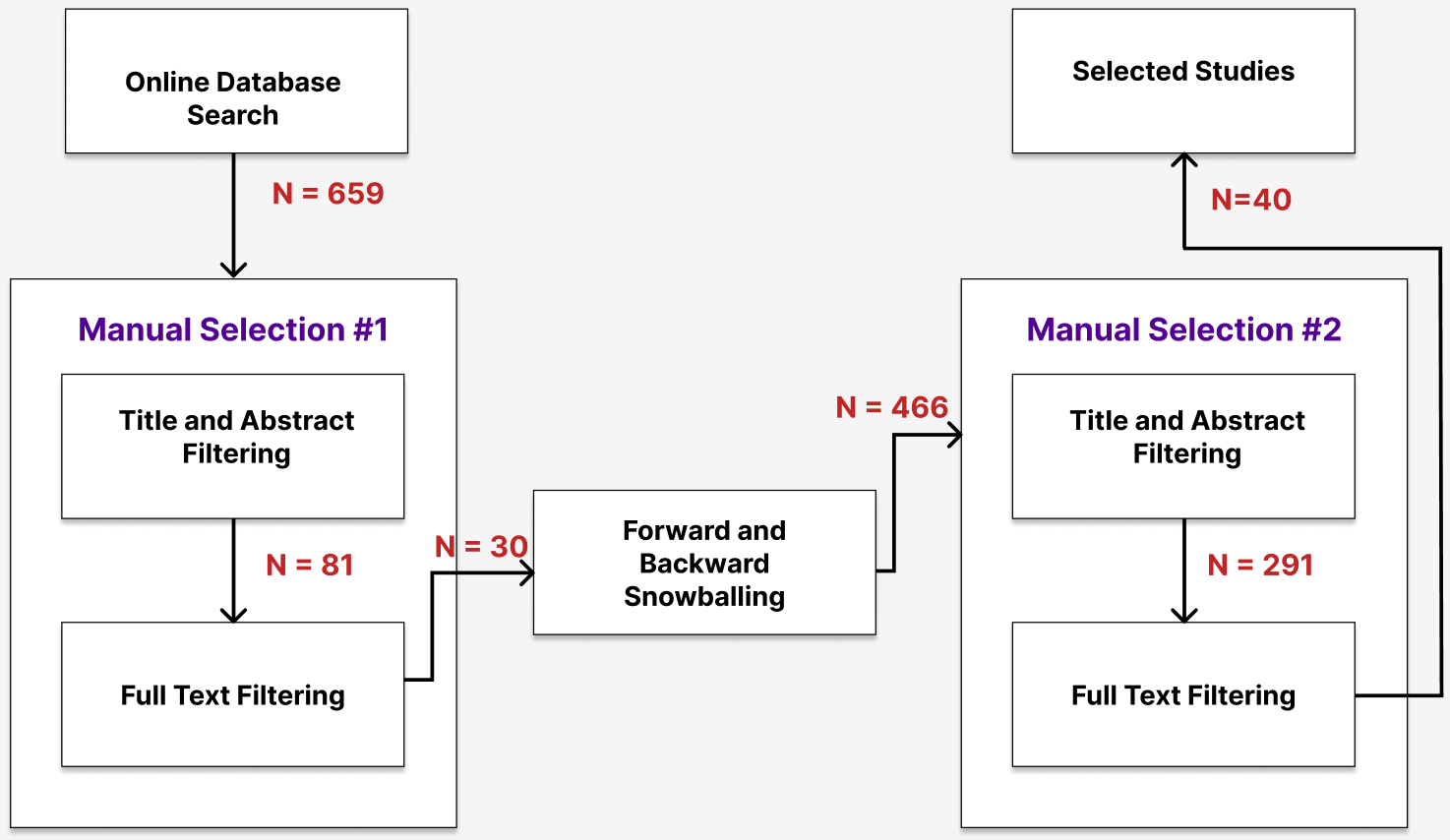}
  \caption{The process of relevant study identification.}
  \label{fig:search_strategy}
\end{figure}

The goal of this paper is to review and analyze the software engineering research that uses LIWC. Following the review study methodology by Kitchenham et al.~\cite{kitchenham2009systematic}, this section outlines our approach toward identifying relevant research papers. An overview of our approach for selecting the relevant papers is shown in Figure \ref{fig:search_strategy}. Additionally, we address PRISMA items in Kitchenham et al.'s checklist for secondary studies to ensure alignment with the systematic review standards~\cite{page2021prisma}. For further reference, please find our PRISMA 2020 checklist attached in the Appendix \ref{app_b}.

To select the papers for our survey, we first  curated a query to retrieve as many relevant research papers in digital libraries as possible. We examined the retrieved papers closely (first title and abstract and then the full text) to filter out any irrelevant or inappropriate studies based on a set of specific inclusion and exclusion criteria (detailed description in Table \ref{tab:ic_ec}). This step was followed by forward and backward snowballing in order to locate any papers that our initial query may have missed. Specifically, we adopted a hybrid approach that combined comprehensive keyword-based searches with both forward and backward snowballing, as supported by recent SLR methodologies \cite{batool2025ai, lin2022opinion, sanchez2019taking}. Finally, the newly collected papers were again manually filtered, leaving us with the final set of papers that we analyzed in this study (n=43). The remainder of this section explains each step of this process in more detail.

\subsection{Online Query} \label{query}

To ensure a comprehensive retrieval of relevant studies, we refined our search query iteratively through a process of trial and validation. Initially, we started with broad queries such as ``liwc" OR "linguistic inquiry and word count", which retrieved a large number of papers across multiple digital libraries, but many were unrelated to software engineering. We then progressively refined the query by adding domain-specific terms, such as ``software", ``software engineering", and ``developer" OR ``development", to filter studies focused on software engineering applications. For instance, adding ``software engineering" significantly reduced irrelevant results while maintaining coverage of key studies. Additionally, we explored broader psycholinguistic and behavioral software engineering terms, such as ``psycholinguistic" OR "psychometric" AND "text analysis" AND "software engineering", to identify studies that might not explicitly reference LIWC but were relevant to our research scope. The final query (``liwc" OR "linguistic inquiry and word count") AND ("software engineering") retrieved the highest number of relevant papers possible. A full record of all queries, retrieval counts, and query refinements is included in our replication package for transparency and reproducibility.

This process, coupled with insights from prior literature regarding suitable phrases for the retrieval of the software engineering papers \cite{lin2022opinion}, led us to select the following search query:

\vspace{.3cm}
\textbf{(``liwc" OR ``linguistic inquiry and word count") AND (``software engineering")}
\vspace{.3cm}

The initial part of the query, \textit{(``liwc" OR ``linguistic inquiry and word count")}, is designed with the intent of inclusively capturing papers that reference the complete name of the targeted tool or its widely recognized abbreviation, ``LIWC". The latter component of our query, \textit{(``software engineering")}, serves as a filter, enabling the identification of papers discussing subjects related to the field of software engineering while not posing as much limitations as terms such as ``developer".

We selected prominent digital archives for our study:  IEEE Xplore Digital Library \cite{ieeexplore}, ACM Digital Library \cite{acmdigitallibrary}, Elsevier ScienceDirect \cite{sciencedirect}, Springer Link Online Library \cite{springerlink}, Wiley Online Library \cite{wiley}, and Scopus \cite{scopus}. The questionable quality of some content on Google Scholar led us to the decision of excluding it as a resource \cite{halevi2017suitability}. Table \ref{tab:retrieved_papers} illustrates the number of papers retrieved from each source in our study.

\begin{table}[htbp]  
  \scriptsize  
  \caption{Number of Retrieved Papers From Each Source}
  \centering
  \begin{tabular}{p{0.5\textwidth} c} 
    \hline
    \textbf{Source} & \textbf{Number of Retrieved Papers} \\
    \hline
    IEEE Xplore Digital Library & 6\\
    ACM Digital Library & 88\\
    Springer Link Online Library & 444\\
    Elsevier ScienceDirect & 56\\
    Wiley Online Library & 8\\
    Scopus & 279\\
    \hline
    \textbf{Total (after duplicate removal)} & 788\\
    \hline
  \end{tabular}
  \label{tab:retrieved_papers}
\end{table}

After eliminating duplicate papers from various sources, we were left with a total of 788 unique documents. Given both the reasonable number of papers and our resource constraints, we decided not to impose any further restrictions on the searches carried out within these online libraries. Our approach involved conducting the searches through the entire body of text rather than limiting ourselves to titles and abstracts alone. Although this method may have led to a higher number of unrelated papers in our results, we deemed it essential to ensure that we identified all relevant papers. This choice was made because not all studies that utilize LIWC at some point in their process explicitly mention it in their titles or abstracts. Next, all 659 articles were subjected to a thorough manual examination, following the method detailed in Section \ref{manual_selection}, to pinpoint the relevant studies for this research.

\subsection{Paper Selection and Eligibility}\label{manual_selection}
To support the manual paper selection process, we first established a precise set of inclusion and exclusion criteria in Table~\ref{tab:ic_ec}, following the PRISMA framework's requirements \cite{page2021prisma}. All the papers included in this study must have undergone peer review and have been published in conferences, workshops, or journals. These papers primarily focus on topics related to software development and use LIWC in their analyses. Additionally, two authors contributed to this manual selection process. The first author used the inclusion and exclusion criteria to identify relevant papers. After each of the two review phases, the second author independently examined the final list, verifying whether each paper’s inclusion aligned with the criteria. Discrepancies or uncertainties were discussed between the two authors until full agreement was reached. This step added a layer of rigor and helped ensure a consistent application of our selection criteria.

\AtBeginEnvironment{tabular}{\scriptsize} 
\begin{table*}  
\caption{Inclusion and Exclusion Criteria}
\begin{tabular}{p{0.05\textwidth} p{0.85\textwidth}}
\hline

     & \textbf{Inclusion Criteria} \\
     \hline \\[1ex]
    IC1 & The paper must have undergone the peer-review process and published at conferences, workshops, or journals
    \\[3ex]
    IC2 & The paper must be available in one of the online libraries mentioned in Section \ref{query}
    \\[3ex]
    IC3 & The conducted study in the paper must contribute to our understanding of the software development process or propose a solution to further advance the related practices.
    \\[3ex]
    IC4 & The study must adopt LIWC to analyze textual data. \\[5ex]
    \hline
     & \textbf{Exclusion Criteria} \\
    \hline \\[1ex]
    EC1 & The paper is not entirely written in English, the dominantly used language for research publications\\[3ex]
    EC2 & The adopted lexicons and methods in the research are specific to a language other than English, which makes the technique not comparable to others.
    \\[3ex]
    EC3 & The paper either serves as an extension of a previously accounted-for conference publication or its extended version has already been included.
    \\[3ex]
    EC4 & The paper is not a traditional research publication and instead serves as an abstract-only submission, doctoral thesis, book, poster, etc.
    \\[3ex]
    EC5 & The paper does not explain what approach was used and in what way LIWC was used.
    \\[3ex]
    EC6 & Papers that do not meet the inclusion criteria of contributing to SE and using LIWC. \\[3ex]
    \hline
  \end{tabular}
  \centering
\label{tab:ic_ec}
\end{table*}

Our manual selection process began by inspecting the titles and abstracts of the papers in order to eliminate irrelevant ones, such as those unrelated to software engineering research, guided by our exclusion criteria. In cases where the information in the title and abstract did not provide sufficient grounds for exclusion, the paper was kept to undergo the next step of our manual selection.

Close examination of titles and abstracts resulted in a collection of 91 potential papers. Subsequently, we examined the text of these 91 papers in their entirety, applying both inclusion and exclusion criteria to determine their eligibility for inclusion in our study. At this stage, we also took note of any information necessary to answer the research questions, e.g., the data types, the data sets and the LIWC categories used in the study. The full-text assessment led to the retention of 36 relevant studies.

As a measure to mitigate any potential limitations of our initial online digital library search, we adopted a forward and backward snowballing approach on the 36 selected papers. This involved gathering all the citations and references cited within these 36 papers, thereby ensuring the inclusion of additional potential papers that may not have been found in our initial search results.

Our snowballing process led to the identification of 494 new and unique papers, which we then subjected to manual selection. Based on the titles and abstracts, we narrowed the number down to 301 potential papers. After conducting a thorough full-text-based filtering, we finally identified 7 additional papers. 

In total, our paper selection process led to the identification of 43 studies (36 in the first iteration + 7 through snowballing) that are the subjects of this survey. The complete set of final papers used throughout this work is listed in Appendix A, where references to each study are provided. Henceforth, we refer to these papers with ``S" prefixes e.g., [S1], [S2], etc. Table \ref{tab:time} illustrates the distribution of our chosen papers based on their publication years. The first software engineering paper using LIWC was published in 2007 [S27]. Since then several software engineering studies have used LIWC to directly analyze many facets of developers' communications. After 2013, the number of software engineering publications that leveraged LIWC in their studies has noticeably grown. As of 2023, LIWC continues to feature prominently in software engineering studies, highlighting the community's ongoing interest and the relevance of this tool in the research area.

\begin{table}[htbp]  
  \scriptsize  
  \caption{The Use of LIWC in Software Engineering Research Over Time}
  \centering
  \begin{tabular}{p{0.25\textwidth} p{0.3\textwidth} p{0.18\textwidth}}
    \hline
    \textbf{Year} & \textbf{Number of Papers} & \textbf{Running Total} \\
    \hline
    2007 & 1 & 1\\
    2012 & 3 & 4\\
    2013 & 5 & 9\\
    2014 & 5 & 14\\
    2016 & 5 & 18\\
    2017 & 4 & 23\\
    2018 & 3 & 26\\
    2019 & 1 & 27\\
    2020 & 4 & 31\\
    2021 & 4 & 35\\
    2022 & 4 & 39\\
    2023 & 3 & 42\\
    Jan-Aug 2024 & 1 & 43\\
    \hline
    \textbf{Total} & \textbf{43} & \textbf{43} \\
    \hline
  \end{tabular}
  \label{tab:time}
\end{table}

\subsection{Data Extraction and Coding}
\label{data_extraction}
In this section, we describe the approach used to extract and code data from each paper included in our review. After retrieving the papers from the sources discussed above, we automatically collected basic metadata for each paper, including its title, author(s), publication year, and venue i.e., journal or conference. This information, along with the papers themselves, is made available in our replication package \footnote{\url{https://figshare.com/articles/dataset/Psycholinguistic_Analyses_in_Software_Engineering_Text_A_Systematic_Literature_Review/28774319}}.

In line with SLR requirements and in order to systematically analyze the content of each paper, we created a data extraction form relevant to our research questions. We used this form to record data such as the LIWC categories utilized, the datasets analyzed, the specific software engineering activities addressed and the behavioral software engineering concepts investigated. The list of all data fields required is presented below:

\vspace{1.5em}
\begin{enumerate}[nosep]
    \item Paper title
    \item Authors
    \item Year of publication
    \item Publication venue (journal/conference)
    \item Purpose for using LIWC
    \item LIWC utilization: direct interpretation or indirect use
    \item LIWC categories used
    \item Dataset(s) analyzed
    \item Software engineering activities studied
    \item Behavioral software engineering concepts addressed
    \item Evaluation methods used to assess LIWC’s effectiveness
    \item Mentioned limitations or concerns with using LIWC
\end{enumerate}
\vspace{1.5em}

For identifying the software engineering activities and behavioral software engineering (BSE) concepts, we adopted a deductive coding approach \cite{thomas2006general}, applying established taxonomies as discussed in Sections \ref{rq2} and \ref{rq3}. For other data extraction tasks, such as identifying the purpose of using LIWC, types of data analyzed, and concerns reported by authors, we employed an inductive coding approach \cite{thomas2006general}, developing categories iteratively based on recurring patterns in the reviewed papers. Two authors independently coded all 43 studies. Following Kitchenham et al.'s guidelines for systematic reviews \cite{kitchenham2009systematic}, we calculated Cohen’s Kappa coefficients for each coding dimension to assess inter-rater reliability. The resulting Cohen’s Kappa coefficients were: 0.61 for both software engineering activities and BSE concepts, and 1.00 for determining the direct/indirect use-cases of LIWC, the purpose for using LIWC, datasets used, evaluation techniques, and concerns or limitations. Where initial agreement was below the threshold of substantial agreement (i.e., less than 0.6), the coding was refined through discussion until substantial agreement was achieved across all dimensions, following the guidelines proposed by Landis and Koch~\cite{Landis77}. The slightly lower agreement for SE activities and BSE concepts can be attributed to the inherent complexity and occasional overlap of categories and labels within each of these dimensions' taxonomies, which may lead to subtle differences in interpretation. Nonetheless, the final levels of agreement, each above 0.6, meet the standard for substantial inter-rater reliability.

\section{Results}
\label{results}

\subsection{RQ1: How was LIWC employed in SE studies, and for what specific activities and purposes?}
\label{RQ1}

\subsubsection{Software Engineering Activities}
We examine specific software engineering activities that were studied using LIWC in SE studies. Borrowing Lin et al.'s taxonomy \cite{lin2022opinion}, we categorized the studies based on the following development activities: \textit{Design Definition Process, Knowledge Management Process, Quality Assurance Process, Stakeholder Needs and Requirements Definition Process, Team Management}. Most papers tackled subjects related to the \textit{Team Management} in development teams. Despite the heavy focus on the human interactions of developers and team members, the studies had different focus areas. For instance, three of the papers that used LIWC to process app reviews focus on \textit{Quality Assurance} activities [S22, S38, S40] and three other studies focus on the \textit{Knowledge Management Process} [S1, S15, S39]. Furthermore, Hellman et al.'s work on analyzing the language used in user forums of projects such as Zotero and Audacity examines the behaviors and expectations of the end-user and therefore makes a contribution to the \textit{Stakeholder Needs and Requirements Definition Process} [S21].

It does not come as a surprise that LIWC, a tool frequently used for automated psycholinguistic analysis, is being used dominantly to better manage developer teams. The more creative approaches can link LIWC's results to the more technical aspect of software development; however, tangible improvements to the software engineering process are not only possible through technical advances \cite{lenberg2015human}. Additionally, many studies point to the existence of correlations between team members' behaviors and projects' outcomes which can be essential to the improvement of the software engineering processes \cite{destefanis2016software, ortu2016emotional, souza2017sentiment}.

\subsubsection{Direct and Indirect Use of LIWC}
\label{direct_indirect_use}
As part of answering RQ1, we examined how LIWC has been used across SE studies and identified two overall usage patterns: \textit{direct} and \textit{indirect}. In Table \ref{tab:applications}, we list all the direct and indirect uses of LIWC along with the categories used for each performed activity and task. We elaborate on both usage types below.

Among all the included studies in our survey, 19 used LIWC to directly analyze text. As the name suggests, direct use of LIWC refers to the cases in which researchers used LIWC's outputs directly to interpret and gain insights from the textual data without any further processing. These use-cases frequently encompass tasks such as detecting developer emotions (e.g., anger, negative emotion, and anxiety) [S13, S18, S28, S30] using LIWC categories like  \textit{affect}, \textit{emotions}, etc. Other tasks included assessing characteristics, such as goal orientation and sociability [S5, S10, S23, S26] and evaluating the linguistic features of LLM responses to Stack Overflow questions.

On the other hand, 21 studies employed LIWC's output indirectly, using it to develop other algorithms or models, or to extract insights that necessitated further processing of LIWC's outputs. The most common method of using LIWC indirectly was to uncover developers' personalities [S2, S8, S9, S19, S27, S35]. Indeed, prior research has demonstrated multiple methods of inferring text authors' personalities using LIWC. Pennebaker \& King \cite{pennebaker1999linguistic}, Yarkoni et al. \cite{yarkoni2010personality}, and Goldbeck et al. \cite{golbeck2011predicting} all have proposed methods of leveraging LIWC for measuring the personality traits of an author based on the widely used Big-Five Factor personality model \cite{goldberg1990alternative}. However, it should be noted that while the original studies that employ LIWC categories to measure personality traits have received significant attention \cite{pennebaker1999linguistic, yarkoni2010personality, golbeck2011predicting}, their focus has predominantly centered on general rather than domain-specific text, such as software engineering-related content.

\begin{longtable}{T{2cm}T{2.6cm}T{.7cm}T{5.9cm}T{1cm}}
\caption{The Direct (D) and Indirect (I) Use of LIWC and The Specific Categories Used for Text Analysis.}\\

    \hline
    \textbf{Activity} & \textbf{Task} & \textbf{D/I} & \textbf{LIWC Categories} & \textbf{Study} \\ \hline
    \endfirsthead
    
    \hline
    \textbf{Activity} & \textbf{Task} & \textbf{D/I} & \textbf{LIWC Categories} & \textbf{Study} \\ \hline
    \endhead
    \hline
    \multicolumn{5}{r}{Continued on the next page} \\
    \endfoot

    \endlastfoot
    
        Team Management & Group Communication Analysis in Teams & D & Pronouns, Cognitive, Work and Achievement, Leisure, Social and Positive, Negative Language & S26, S31, S32, S34, S36 \\
    
         & Group Communication Analysis in Teams & D & Cognitive, Work and Achievement, Leisure, Social and Positive, Negative Language & S29, S33 \\ 

         & Identifying Developer Preferences & I & Visual, Auditive, ``Kinaesthetic" & S25 \\ 
        
        & Group Communication Analysis in Teams & D & Achieve, Affect, Discrep, Funct, Hear, Humans, Leisure, Negative Emotion, Numerals, Past, Present, QuestionMark, Social  & S18 \\
        
        & Group Communication Analysis in Teams & D & Work and achievement, negative, cognitive, social and positive & S17 \\

        & Group Communication Analysis in Teams & D & Linguistic process, Linguistic categories, Psychological Processes, relativity, current concerns, spoken categories, and punctuation & S6 \\ 
        
        & Group Communication Analysis in Teams & D & Pronouns, Cognitive, Work and Achievement, Emotions & S5 \\
        
        & Group Communication Analysis in Teams & D & Pronouns, Work and Achievement, Cognitive, Leisure, Social and Emotions & S10 \\ 

        & Group Communication Analysis in Teams & D & Achievement, Work, Positive, Negative, Insight, Social, prop(I) & S37 \\ 

        & Group Communication Analysis in Teams & D & prop(focusfuture), prop(we), prop(posemo), prop(negemo), prop(certain), prop(tentative) & S9 \\ 

        & Group Communication Analysis in Teams & I & \textit{Categories Not Specified} & S4\\ 

        & Group Communication Analysis in Teams & I & Personal pronouns, Impersonal Pronouns, Articles, Prepositions, Auxiliary verbs, Common adverbs, Conjunctions, Negations, Analytical Thinking, Clout, Authentic, and Emotional Tone. & S24 \\ 

        & Group Communication Analysis in Teams & D \& I & Standard WC Pronouns, Cognitive, Relative, Optimism, References to Time, Future tense verbs, Social Processes, Categories correlated with personality traits in \cite{pennebaker1999linguistic} & S27\\ 
        
        & Personality Detection & I & Categories correlated with personality traits in \cite{pennebaker1999linguistic} & S2, S20, S23 \\
        
        & Personality Detection & I & Categories correlated with personality traits in \cite{yarkoni2010personality} and \cite{golbeck2011predicting}  & S19 \\ 
        
        & Personality Detection & I & Categories correlated with personality traits in \cite{yarkoni2010personality} & S7 \\
        
        & Personality Detection & I & Categories correlated with personality traits in \cite{yarkoni2010personality}, Categories used in Personality Recognizer tool \footnote{http://s3.amazonaws.com/mairesse/research/personality/recognizer.html} & S8 \\ 
        
        & Personality Detection & D \& I & Categories correlated with personality traits in \cite{pennebaker1999linguistic}, Cognitive, Work and Achievement, Leisure, Social and Positive, Negative Language & S35 \\ 

        & Emotion Detection & D & Positive affect (Positive Emotion), Negative affect (Negative Emotion, Sadness, Anger, Anxiety, Inhibition) & S30 \\ 
        
        & Emotion Detection & D & Anger & S28 \\
        
        & Emotion Detection & I & \textit{Categories Not Specified} & S12, S13 \\ 
        
        & Emotion Detection & I & Anger & S14 \\ 

        & Sentiment Analysis & I & \textit{Categories Not Specified} & S3 \\ 
        
        & Argument Extraction & D \& I & \textit{Categories Not Specified} & S11 \\ 

        \hline
    
        Quality Assurance & Predicting Deletion of Stack Overflow Post & I & Personal Pronouns, Pronouns, Space, Relativity, Inclusive, Cognitive Process, Social, 1st Person Singular Pronouns, Function Words, Conjunctions, Prepositions & S1, S39 \\

        & LLM Evaluation & D & Affective Process, Cognitive Process, Perceptual Process, Informal language & S43 \\

        & Insider Threat Detection & I & \textit{Categories Not Specified} & S41 \\

        \hline

        Stakeholder Needs and Requirements Definition & User Input Analysis & D & Clout, Authentic, Analytic, Tone, Cognitive process, Affective process, Drives & S21 \\ 

        & User Input Analysis & I & Negative Emotion & S15, S22, S38, S40 \\

        \hline
        
        Design Definition Process & Design Bias Analysis & D & Personal Concerns (“persconc”), Social and Affect & S42 \\ 

    \hline \\

\label{tab:applications}
\end{longtable}

\subsubsection{Notable Applications of LIWC}
Through answering RQ1, we also aim to better understand the types of LIWC applications within the SE research. We pinned down the purposes LIWC has fulfilled in each study. Here, in more detail, we discuss some notable applications of LIWC in SE research.

To predict the likelihood of a post's deletion on Stack Overflow, researchers have used LIWC to analyze its content. Given that many Stack Overflow posts can be either off-topic or low quality (predominant reasons of deletion~\cite{duijn2015quality}), the task of manually sifting through the posts and deleting such content is laborious and time-consuming. Consequently, the development of automatic methods for identifying posts that need to be deleted can offer significant support to those responsible for monitoring the website's content. Correa \& Sureka created a classification model trained on 47 features, some of which were created using the textual contents of the posts. Within these features were 11 LIWC categories such as \textit{Relativity}, \textit{Cognitive}, \textit{Inclusive}, and \textit{Social} that were considered potentially distinctive for this task. Using these features, a prediction model with 66\% accuracy was developed. Additionally, this study found new insights into the quality of the user-generated content and characteristics of the deleted posts on Stack Overflow. This study was later built upon and improved in another work where the 47 features were reused in a different model to improve the prediction of Stack Overflow posts that need to be deleted [S39].

LIWC has also been used in the development of other psychometric tools. In another work, Santos et al. present \textit{NEUROMINER}, a psychometric tool based on LIWC, that identifies developers' preferred representational systems [S25]. Representational systems refer to the ways in which the human brain stores and processes information. Often within the NLP domain, these systems are recognized as \textit{Visual}, \textit{Auditive}, or \textit{Kinaesthetic} \cite{druckman1988enhancing}. Knowing the developers' preferred representational systems can help us better understand the most effective methods of communication and knowledge sharing. \textit{NEUROMINER} extracts LIWC values from an input text, and performs ANOVA  (analysis of variance) to detect the preferred representational systems of the top four Apache committers. The authors use \textit{Visual}, \textit{Auditive}, and \textit{Kinaesthetic} categories of LIWC to decide which of these representational models a developer tends to prefer.

Zolduoarrati et al. studied the impact of cultural differences in developer communications with the help of LIWC. The authors conducted a study, examining Stack Overflow users' behaviors across three distinct countries, characterized by varying degrees of individualistic or collectivist values [S10]. The culture of the selected countries i.e., the United States, China, and Russia, exhibit different levels of individualism, as determined by Hofstede's cultural index \cite{hofstede1984cultural}. The authors argue that understanding the cultural patterns of users from these countries holds potential for improving communication and cooperation in globally distributed software development efforts. To accomplish this, the study leveraged LIWC categories, including \textit{Cognitive}, \textit{Emotional}, and \textit{Leisure} dimensions, to analyze the text written by the users from these countries.

Furthermore, researchers have used LIWC to examine the topic of trust among developers in a collaborative team. Even though some studies on this topic exist \cite{sajadi2023interpersonal, lavazza2010predicting, calefato2017preliminary}, few  attempted to automatically measure trust \cite{da2018arsenal, venigalla2021understanding}. Wang \& Redmiles have, however, combined NRC Lexicon \cite{mohammad2013crowdsourcing} and some dimensions of LIWC related to trust to measure a baseline of trust among developers [S4]. They did not specify the exact LIWC categories used; nevertheless, they evaluated their findings by calculating the baseline level of trust with NLP methods based on the communications of 10 individuals involved in their studied projects.

Janjua et al.'s study addresses the growing challenge of insider threats in information technology systems, where insiders can bypass security measures and access confidential documents [S41]. In their work, they utilized the content of email communications and the output from LIWC analysis to train various machine learning models aimed at detecting insider threats.

Also worth mentioning is Han et al.'s study on the cross-status communications of the developers in open source software development [S24]. In their work, the authors use 12 LIWC categories to compute the similarities of the language styles used by different groups of developers and explore the possibility of a correlation between the project outcomes and the cross-status communications between \textit{elite} and the \textit{non-elite} developers.

Additionally, [S19] and [S8] examine LIWC's ability to infer developers' personalities and compare the tool with the now-discontinued LIWC-based tool, IBM Personality Insights \footnote{https://watson-developer-cloud.github.io/node-sdk/master/classes/personalityinsightsv3.html}. While [S19] finds the two tools to produce consistent results, [S8] points out a noticeable level of disagreement between LIWC and IBM Personality Insights.

Finally, in attempting to analyze the opinions of the community in open source software development, LIWC has been utilized to create the Argulens framework [S11]. Argulens creators explore how usability discussions occur in issue tracking systems, and how effective the machine learning models can be in extracting arguments. They also discuss the potential of the framework to support practitioners in understanding community opinions. LIWC's output is used alongside other features as an input to the machine learning classifiers for tasks such as classifying the Argumentative vs. Non-argumentative conversations.

\begin{mdframed}[backgroundcolor=lightgray!20,topline=false,leftline=false,rightline=false,bottomline=false] 
\textbf{RQ1 Summary}: 
LIWC has been directly and indirectly utilized in a variety of SE tasks to understand the nuances of developer learning, collaboration, and communication. The most common uses include personality detection and emotion analysis. Other applications include quality assessment of forum posts, investigation of cultural differences in distributed teams, and argument extraction.

\end{mdframed}

\subsection{RQ2: What types of textual data were analyzed using LIWC in SE studies?}
\label{rq2}
\begin{table}[htbp] \scriptsize
  \caption{Data Types and Sources}
  \begin{tabularx}{\linewidth}{XXX}
    \hline
    \textbf{Data Type} & \textbf{Sources} & \textbf{Studies} \\
    \hline
    Interactions on project management platforms & IBM rational jazz & S2, S17, S26, S29, S31, S32, S33, S34, S36, S37\\
    & GitHub issues & S4, S7, S9, S11, S14, S19, S24, S28 \\
    && \\
    & GitHub pull requests & S7, S9, S19, S24, S28\\
    && \\
    &GitHub code reviews & S7, S24, , S9, S19 \\
    && \\
    &Jira issues& S12, S13 \\
    && \\
    &GitHub discussions & S4 \\ 
    \hline
    Q\&A & Stack Overflow& S1, S3, S5, S10, S20, S23, S35, S39\\
    &&\\
    & Kaggle forums & S6 \\
    &&\\
    & ChatGPT answers to Stack Overflow questions & S43 \\
    && \\
    & Specific software product forums (VLC, Zotero, etc.)& S21 \\
    \hline
    Email Communications & Apache Software Foundation mailing lists & S4, S8, S16, S25, S27\\
    & Malicious users' interactions & S41\\
    \hline
    User Feedback & App Reviews & S15, S22, S38, S40\\
    \hline
    Chats & IRC & S4\\
    &&\\
          & Controlled Studies & S18 \\
    \hline
    Microblogs & OfficeTalk software (developed exclusively for internal use in Microsoft) & S30 \\
    \hline
    Written Solutions to Design Tasks & Conducted Experiments & S42 \\
    \hline
  \end{tabularx}
  \label{tab:data}
\end{table}

To categorize the software engineering texts analyzed with LIWC, we identified the types of developer communication used in each study. Table \ref{tab:data} shows the data types and sources of datasets used in the 43 papers we examined.

\begin{figure}[htbp]
    \centering
    \includegraphics[width=\linewidth]{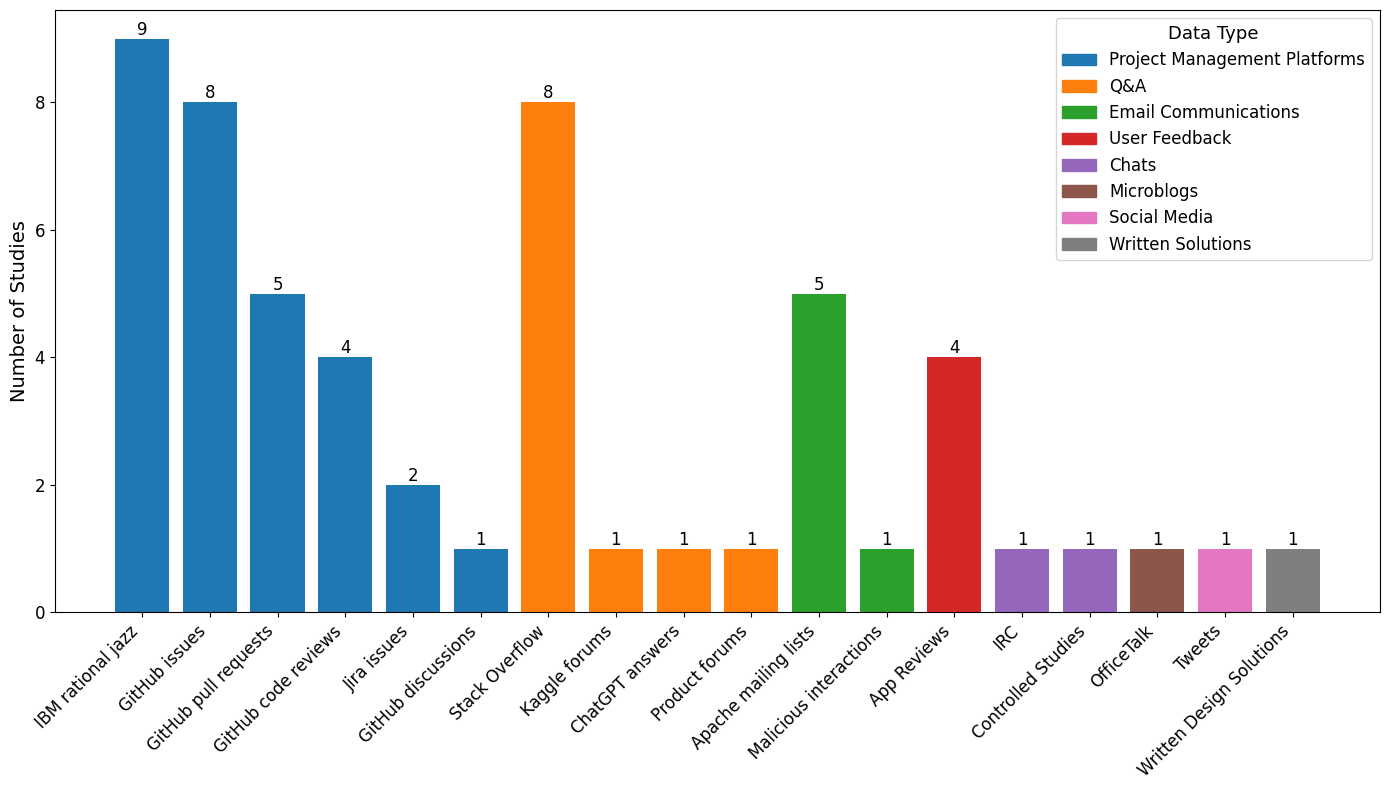}
    \caption{Distribution of Data Sources Used in SE Studies Employing LIWC. The figure highlights the prevalence of project management platforms and Q\&A forums as dominant sources of analyzed text.}
    \label{fig:data_sources}
\end{figure}

Most researchers focused on analyzing the communications of developers on project management platforms, i.e., IBM Rational Jazz, GitHub, and Jira. In 10 of the 43 studies, IBM Rational Jazz was used as the primary source of data. In addition to hosting developers' communications, IBM Rational Jazz provides environments for both development and tracking the progress of the project. These 10 papers use LIWC to study developers' comments in Jazz repositories over time, and to explore various aspects of developer interactions and social dynamics, including the impact of leadership and knowledge sharing on project success [S26, S31]. Figure \ref{fig:data_sources} provides a visual overview of the distribution of data sources across the studies we reviewed.

Q\&As are another type of commonly used data that is analyzed with the help of LIWC. Of the 11 studies that have focused on Q\&A forums [S6, S21], the most popular Q\&A platform, Stack Overflow, has been used in 8 different studies. While some studies using Stack Overflow data focus on personality detection [S20, S23, S35], other studies use the data to identify low quality posts [S1, S39], different attitudes of developers based on their cultural background [S10], developer sentiment [S3], and in one case, exploring differences in the attitudes of male and female developers [S5]. Additionally, one study used Stack Overflow questions as prompt for ChatGPT, in order to compare the LLM-generated answers to those provided by users [S43].

The Apache Software Foundation (ASF) mailing lists have also served as data sources for a number of studies.  ASF mandates that projects must maintain full transparency in their communications through public mailing lists, following the commonly held principle that, if it wasn't documented on the mailing lists, it didn't happen~\cite{Likang2021, 4228660}. By requiring all the developer communications to take place on mailing lists, ASF creates a valuable source of data for those interested in analyzing the interactions of many ASF projects' team members. Such analyses were used to study the trust dynamics among the developers, developers' behaviors, and developers' persoanlities [S4, S8, S27].

Four studies used LIWC to analyze app reviews [S15, S22, S38, S40]. All 4 of these studies used LIWC specifically to identify the words associated with the negative emotions according to LIWC's lexicon. The findings were then combined with other observations and data points extracted from the reviews. Authors of these 4 papers use the findings to suggest prioritizing the improvement of certain app features over others, draw conclusions about user satisfaction with an app, etc.

 Notably, a few studies used LIWC to analyze software engineering chats and microblogs [S4, S18, S30]. Given the prevalence of developers' emotional expressions in chats \cite{sajadi2023towards, chatterjee2019exploratory, subash2022disco}, exploring various emotional and psychological factors through developers' chat communications may yield novel insights into the human factors affecting the development process.

\begin{mdframed}[backgroundcolor=lightgray!20,topline=false,leftline=false,rightline=false,bottomline=false] 
\textbf{RQ2 Summary}: The most frequent data sources include project management platforms (e.g., GitHub) and Q\&A forums (e.g., Stack Overflow). Other data sources include mailing lists, app reviews, and developer chats. This diverse set of data sources highlights LIWC’s flexibility in processing different forms of communication in the SE domain.
\end{mdframed}

\subsection{RQ3: What Behavioral Software Engineering (BSE) concepts were studied using LIWC?}
\label{rq3}
Behavioral Software Engineering (BSE) is defined as ``the study of cognitive, behavioral and social aspects of software engineering performed by individuals, groups or organizations.''\cite{lenberg2015behavioral}. BSE combines social sciences and empirical software engineering to gain insight into human factors in software engineering \cite{felipe2023psychometric, lenberg2015behavioral, feldt2008towards}. Through their work, including studying psychology textbooks, examining related papers, and interviews with field experts, Lenberg et al. identified 55 core concepts underpinning BSE research, such as \textit{Group Norms}, \textit{Organizational Climate}, and \textit{Leadership} \cite{lenberg2015behavioral}. 

Understanding how LIWC has been used to study BSE concepts reveals the tool's  effectiveness in capturing critical human factors that directly influence software engineering practices and outcomes. Answering this research question contributes to  assessing LIWC's role in advancing behavioral research within SE. Next we discuss in detail the studied behavioral software engineering concepts using LIWC in our 43 selected studies.

To find out whether or not LIWC has been used to conduct software engineering-related studies tackling different BSE concepts, we have identified all the BSE concepts studied in each of our selected studies. Table \ref{tab:bse} illustrates the definitions of the BSE concepts that have been borrowed from ~\cite{lenberg2015behavioral} and identified in one or more of our selected 43 studies.

\begin{longtable}{p{0.15\linewidth}p{0.6\linewidth}p{0.17\linewidth}}
\caption{\scriptsize{BSE Concepts found in our study}}
\label{tab:bse}

 \\ \toprule
    \textbf{BSE Concept} & \textbf{Definition} & \textbf{Studies} \\
    \hline
    \endfirsthead 
    \multicolumn{3}{c}{{\tablename\ \thetable{} -- Continued from previous page}} \\
    \hline
    \textbf{BSE Concept} & \textbf{Definition} & \textbf{Studies} \\
    \hline
    \endhead 
    \hline
    \multicolumn{3}{c}{{Continued on next page}} \\
    \endfoot 
    \hline
    \endlastfoot
  
    \hline
    Communication & ``The process that allows people to exchange information, feeling or thoughts " \cite{lenberg2015behavioral, katz1978social} & All selected studies\\

    Organizational Climate & ``Is defined as the recurring patterns of behavior, attitudes and feelings that characterize life in the organization (Isaksen et al., 2007; Denison, 1996)." \cite{lenberg2015behavioral, isaksen2007assessing, denison1996difference} & S4, S5, S7, S12, S14, S24, S27, S28, S29, S30, S31, S32, S33, S34, S36, S37\\

    Motivation & ``According to Pardee (Pardee, 1990) motivation can be defined as those forces within an individual that push or propel him to satisfy basic needs. It is what prompts a person to act in a certain way or at least develop an inclination for specific behavior." \cite{lenberg2015behavioral, pardee1990motivation} & S7, S9, S12, S17, S21, S22, S26, S31, S33, S36, S37, S38, S40\\
    
    Positive psychology & ``Positive psychology is the branch of psychology that study the strengths and virtues that enable individuals and communities to thrive (Seligman and Csikszentmihalyi, 2000)." \cite{lenberg2015behavioral, seligman2000positive} & [S3, S17, S18, S23, S26, S27, S28, S29, S30, S31, S33, S35, S37\\
    
    Personality & ``According to Feist and Feist (Feist, 1994) personality is a pattern of relatively permanent traits and unique characteristics that give both consistency and individuality to a person’s behavior." \cite{lenberg2015behavioral, feisttheories} & S2, S7, S8, S17, S19, S20, S27, S35, S37 \\
    
    Leadership & ``Is the art of influencing followers to achieve success by identifying joint goals, finding best-fit roles in teams, collaborating constructively and dynamically, and adapting to change within their environments (Wikipedia, 2014)." \cite{lenberg2015behavioral, WikipediaLeadershipPsychology} & S9, S18, S27, S29, S24, S31\\
    
    Group dynamics & ``Is a system of behaviors and psychological processes that occurs within a group or between groups (Forsyth, 2009)." \cite{lenberg2015behavioral, forsyth2018group} & S2, S4, S6, S10, S11, S24, S28\\
    
    Group norms & ``Are informal rules that regulates the behavior of the group (Feldman, 1984)." \cite{lenberg2015behavioral, feldman1984development} & S5, S6, S24, S36\\

    Cognitive style & ``Individual differences regarding strategies for perceiving, remembering, thinking, and problem solving" \cite{lenberg2015behavioral, messick1984nature} & S16, S25\\
    
    Intentions to leave & ``Refers to conscious and deliberate willfulness to leave the organization (Tett and Meyer, 1993)." \cite{lenberg2015behavioral, tett1993job} & S27\\
    
    Organizational Change & ``Is both the process in which an organization changes and the effects of these changes on the organization (Portal, 2014). " \cite{lenberg2015behavioral, StudyComOrganizationalChange} & S27\\
    
    Group social identity & ``Defined by Hogg and Vaughan as an self-concept derived from perceived membership of social groups, i.e., the aspects of a person that are defined in terms of his or her group memberships (Hogg and Vaughan, 2002)." \cite{lenberg2015behavioral, vaughan2013social} & S10\\

    Risk taking & ``Is the tendency to engage in potential harmful behavior that, at the same time, could have some kind of positive outcome (March and Shapira, 1987)." \cite{lenberg2015behavioral, griesinger1973toward} & S41\\
    
    Stress & ``Defined by the American Psychological Association as a transient state of arousal with typically clear onset and offset patterns" \cite{lenberg2015behavioral, APAStress} & S14 \\

\end{longtable}

\textit{Communication}, in particular stands out as the single most studied BSE concept as the researchers employing LIWC consistently analyze various forms of textual communication, whether among developers themselves or between other stakeholders and developers, as a means of discovering new information. 

\textit{Organizational Climate}, \textit{Motivation}, \textit{Positive Psychology}, and \textit{Personality} represent the most extensively explored concepts within the collected body of literature. Among our subjects of study, the least commonly studied BSE concepts were \textit{Risk taking}, \textit{Stress}, and \textit{Group Dynamics}.

\begin{figure}[h]
    \centering
    \includegraphics[width=1\linewidth]{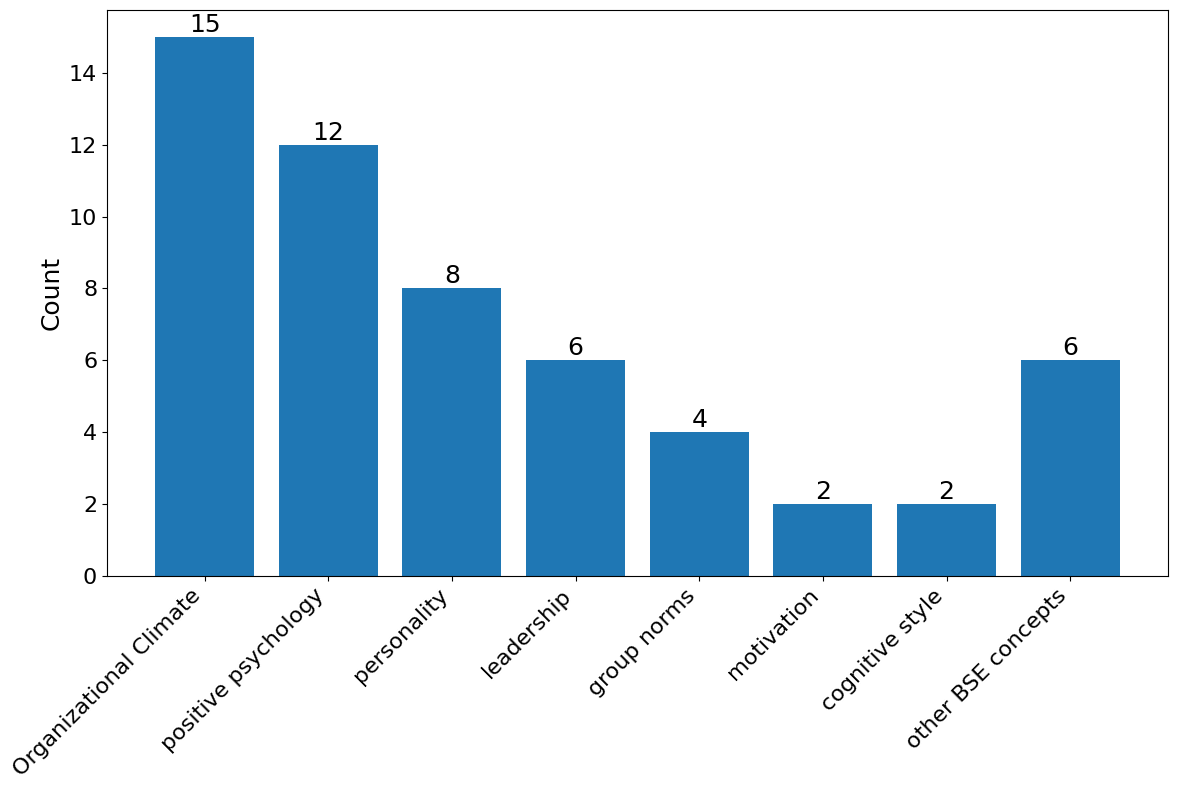}
    \caption{Most Common BSE Concepts Studied in Software Engineering Literature Using LIWC.}
    \label{fig:bse_common}
\end{figure}

In line with Lenberg et al.'s results, indicating \textit{Communication} and \textit{Personality} to be among the most studied BSE concepts, we found both of these concepts to be frequently examined in papers employing LIWC in software engineering. However, contrary to the observations by Lenberg et al., we found that \textit{Organizational Climate} and \textit{Positive Psychology} also to be frequently examined and studied in a number of papers. Figure \ref{fig:bse_common} illustrates the most commonly studied BSE concepts in our selected literature.

While some BSE concepts, such as sentiment or emotion, have long been studied using a range of automated tools, others like \textit{Organizational Climate} have traditionally relied on qualitative methods, such as interviews or surveys, due to their abstract and contextual nature. However, LIWC has enabled researchers to examine these social concepts in a more scalable and automated fashion through text analysis. This may help explain why \textit{Organizational Climate} appears more frequently in our set of studies than reported in earlier BSE reviews \cite{tsai2010programmer}. Similarly, BSE and social concepts like \textit{Stress} and trust, which might otherwise be difficult to quantify or observe without self-reporting, become more accessible through LIWC’s built-in stress-related categories. In this way, LIWC lowers the barrier to analyzing many complex psychological or social phenomena in SE, helping to expand the reach of BSE research.

\begin{mdframed}[backgroundcolor=lightgray!20,topline=false,leftline=false,rightline=false,bottomline=false] 
\textbf{RQ3 Summary}: 
We present all the BSE concepts studied with the use of LIWC in Table \ref{tab:bse}. The most commonly examined concepts include \textit{Communication}, \textit{Motivation}, \textit{Organizational Climate}, \textit{Positive Psychology}, and \textit{Personality}, while concepts such as \textit{Risk Taking}, \textit{Stress}, and \textit{Group Dynamics}, are less commonly studied.

\end{mdframed}

\subsection{RQ4: How often and in what way has LIWC been evaluated in a software engineering context?}

Like any automated analysis tool, LIWC must be evaluated in order for us to determine its effectiveness within the context of SE studies. In the context of our review, evaluation refers to any effort by SE researchers to assess the effectiveness or reliability of LIWC in their studies. Such evaluations could involve expert assessments, for example, a psychologist reviewing whether LIWC is accurately capturing relevant psycholinguistic signals, or comparisons between LIWC output and established ground truth data. To answer RQ4, we identify and categorize the various ways in which software engineering researchers have evaluated the efficacy of LIWC in their studies.

\begin{figure}[htbp]
  \centering
  \includegraphics[width=0.8\linewidth]{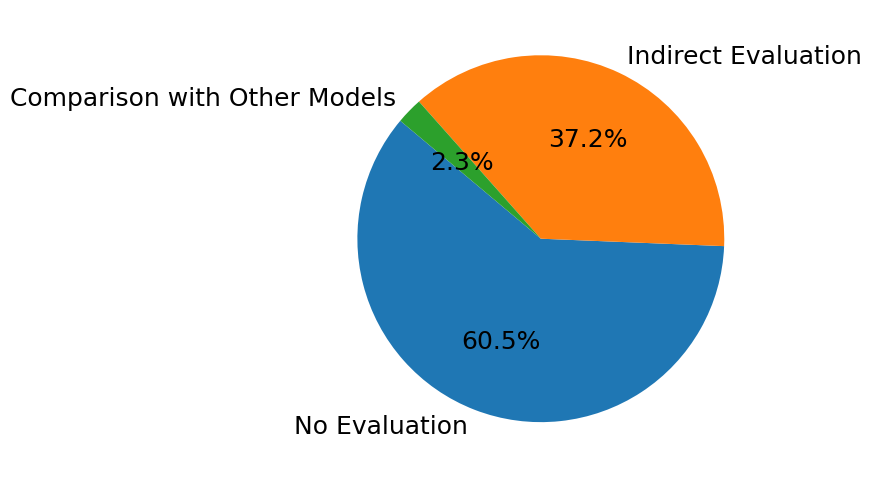}
  \caption{Distribution of evaluation strategies used in SE studies employing LIWC.}
  \label{fig:liwc_eval}
\end{figure}

A notable observation among the papers that used LIWC is that the majority, specifically 26 of 43 papers, did not formally evaluate LIWC's output (see Table \ref{tab:evaluation} and figure \ref{fig:liwc_eval}). These papers relied on LIWC for various purposes and accepted the resulting psychometric measures without any evaluation. In these studies, LIWC's output was either used for qualitatively analyzing the text or was combined with other data to facilitate subsequent analysis. Some of these studies have used the combination of LIWC's output and other data points to train machine learning models for tasks such as the prediction of the low quality Stack Overflow posts [S1, S39] or the identification of different arguments within a discussion i.e, argument extraction [S11]. These papers mainly justified their decision based on the established reputation of LIWC as a widely recognized and reliable psychometric text analysis tool, cited by thousands of studies~ \cite{tausczik2010psychological}.

\AtBeginEnvironment{tabularx}{\scriptsize}
\begin{table}
\caption{List of Papers With and Without Evaluation methods for LIWC's Output}
\vspace{.2cm}
\label{tab:evaluation}
\begin{tabularx}{\linewidth}{XX} 
    \hline 
    \textbf{Evaluation Method} & \textbf{Papers} \\ 
    \hline
    No Evaluation & S2, S5, S6, S7, S10, S11, S16, S17, S18, S20, S21, S22, S23, S24, S26, S28, S31, S32, S33, S34, S35, S36, S40, S41, S42, S43 \\ 
    \hline
    Indirect Evaluation & S1, S3, S4, S8, S12, S13, S14, S15, S19, S25, S27, S29, S30, S37, S38, S39 \\ 
    \hline
    Comparison with Other Models & S9 \\ 
    \hline
\end{tabularx}
\end{table}

Sixteen of the papers evaluated LIWC indirectly, meaning that instead of evaluating the output of LIWC itself, they assessed the tools or models built using LIWC as part of their feature sets. These evaluations often focused on the performance of the models or tools, sometimes incorporating ground-truth data, such as personality questionnaires, to measure accuracy or effectiveness. For example, some studies used LIWC’s categories in machine learning models for tasks like predicting low-quality Stack Overflow posts [S1, S39] or extracting arguments from discussions [S11]. [S38] also combined LIWC’s negative words with other lexicons to create a tool that detects 'expectation violations' in mobile app reviews, yielding an accuracy of 96\% on a manually labeled subset of 400 data points. In these studies, the focus was on evaluating the overall performance of the models or tools rather than directly assessing LIWC's individual output.

One study, [S9], performed a comparison of LIWC’s sentiment analysis performance with another model, NLTK VADER, and found them to be perform comparably. However, this was not a formal evaluation of LIWC itself, as it simply compared the performance between the two tools without directly assessing LIWC’s internal accuracy or outputs in isolation. Despite this, we still wanted to distinguish it as a valuable comparison for understanding how LIWC performs relative to other sentiment analysis tools.

Although LIWC has established a strong reputation and is widely used in the scientific community, there remains a need for more formal evaluations of its effectiveness within the software engineering context. While LIWC’s psychometric measures are often trusted without direct evaluation, formally assessing its accuracy and applicability within SE research would benefit the community. However, this is not always a straightforward task. Evaluating LIWC in tasks like sentiment analysis may be relatively simple, but assessing its measures of psychological concepts, such as personality or emotions, is more complex and requires validation through psychological instruments like specific questionnaires or personality tests. So far, no significant effort has been made to evaluate LIWC directly within software engineering, leaving a gap for future research to address.  

\begin{mdframed}[backgroundcolor=lightgray!20,topline=false,leftline=false,rightline=false,bottomline=false] 
\textbf{RQ4 Summary}: Out of the 43 papers reviewed, 26 did not conduct a formal evaluation of LIWC's output, relying instead on its established reputation in the scientific community. Of the remaining 17 papers, most assessed LIWC indirectly by evaluating models or tools that incorporated LIWC’s outputs, often using ground-truth data.
\end{mdframed}

\subsection{RQ5: What concerns or limitations were mentioned by the researchers adopting LIWC in the software engineering domain?}

\AtBeginEnvironment{table}{\scriptsize}
\begin{table*}[htbp]{\scriptsize}
\centering
  \caption{\scriptsize{LIWC-related Limitations and/or Concerns Expressed in the Selected Studies}}
  \label{tab:all_limitations}
  \footnotesize
  \begin{tabular}{p{0.65\linewidth}p{.3\linewidth}}
    
    \hline
    \textbf{Limitation of LIWC}& \textbf{Studies} \\ 
    \hline 
     No concern or limitation expressed & S1, S3, S4, S5, S6, S8, S11, S12, S13 S14, S15, S16, S18, S20, S22, S23, S24, S25, S28, S30, S33, S35, S36, S37, S39, S40, S41, S42, S43\\
    Domain specificity (e.g., lack or misrepresentation of SE words) & S2, S10, S19, S26, S27, S31, S32, S34, S38 \\
    Language limitation (e.g., English lexicon) &  S7 \\
    Bias against non-native speakers & S27 \\
    Lack of qualitative analysis &  S21 \\
    Limited sentiment analysis capabilities on informal communications & S9 \\
    Inability to account for negation & S29 \\
    \hline  
  \end{tabular}
\end{table*}

To answer this RQ, we have outlined all the concerns or limitations expressed in the 43 papers we studied. Table \ref{tab:all_limitations} presents an overview of how these studies have either acknowledged or overlooked the limitations associated with the use of LIWC. 29 out of the 43 papers refrain to acknowledge potential limitations of using LIWC. In 14 studies, the authors acknowledge the conceivable limitations of LIWC, delving deeper into the issues or offering potential solutions to mitigate the problems to varying degrees. Some of these studies offer detailed explanations of the concerns pertaining to the adoption of LIWC. We find that these concerns are either inherent limitations of lexicon-based tools or otherwise, related to the application of LIWC to software engineering. Next, we discuss both types of concerns.

\subsubsection{Concerns with LIWC's inherent limitations as a lexicon-based tool}
\label{inherent_limitations}

Much like any other lexicon-based tool, the use of LIWC comes with its own set of limitations. Some of these limitations are inherent and often bound by its technical constraints, such as the \textit{inability to account for negation or understanding the context} in a discussion [S29]. For instance, according to LIWC, the sentence ``This is not good" contains positive emotion, since the word ``good" is associated with positive emotion in LIWC's dictionary.

Another notable limitation of LIWC is its \textit{limited availability of dictionary languages}, as highlighted by Rigby et al. [S27]. 
While LIWC2015 includes a German dictionary as its sole non-English option, it's worth noting that individuals have voluntarily undertaken the considerable task of translating the English LIWC dictionaries (2001, 2007, and 2015 versions) into languages such as simplified and traditional Chinese, Spanish, Italian, Russian, etc. These translated dictionaries are accessible to paid users on LIWC's official website\footnote{https://www.liwc.app/dictionaries}. The limited number of translated dictionaries may be attributed to the substantial effort required for translation, thus explaining their scarcity.
Another language-related limitation of LIWC concerns \textit{text written by non-native speakers}. Non-native speakers often use a more restricted vocabulary in their communications, potentially influencing the outcomes generated by LIWC [S7].

A potential concern with psycholinguistic tools is that authors might intentionally manipulate their language to influence the analysis. However, this is largely irrelevant in SE contexts, where communications, such as those on GitHub, are organic and not written with the expectation of being psycholinguistically analyzed [S7]. As such, we do not consider this a meaningful limitation for our review.

Overall, while the criticisms aimed at LIWC's inherent constraints are valid, as it stands, LIWC has been capable of drawing many conclusions from different general-purpose texts successfully \cite{tausczik2010psychological}. Additionally, as highlighted in the past studies [S29], in practice, when analyzing large corpora of text, the sheer abundance of the data can compensate, to a certain extent, for some of the LIWC's shortcomings. Therefore, in order to improve LIWC's performance for software engineering-specific text we might need to turn our attention to the concerns regarding the adoption of this tool in the software engineering domain.

\subsubsection{Concerns with LIWC's adoption in SE}

Many studies acknowledge the limitations of adopting LIWC in a software engineering context. Previous research that has identified numerous correlations between various LIWC categories and a wide range of social and psychological phenomena has primarily focused on general text~\cite{tausczik2010psychological}. Given the complexities associated with the transition of psychometric tools to other fileds \cite{felipe2023psychometric}, care should be taken when adopting tools such as LIWC in Software Engineering studies.

LIWC, may miss many of the words used in particular domains such as software engineering, since \textit{domain specific words may be absent} from its dictionaries. LIWC's output is as good as its dictionaries and therefore, if the dictionary misses many of the words present in the text, the reliability of LIWC's output can be severely questioned. It has been observed that some older versions of LIWC can only capture around 66\% of the words in software engineering text, discarding words such as API, XML, URI, and HTTP [S2, S31, S34, S28]. However, it is worth mentioning that LIWC's developers, have been extending its dictionaries on each new release and although the size of a dictionary will always remain limited, LIWC-22 has the largest LIWC dictionary yet, with a size of over 12000 lexicons.

The \textit{misrepresentation of domain-specific words} can also limit the use of LIWC in software engineering research. In many cases, the performance of the tool can be questioned since certain words can have different meanings in particular contexts compared to general text. Certain software engineering-specific terms like ``bug", ``kill", or ``cookie" may not be accurately represented in a LIWC dictionary. For instance, the term ``cookie," appears in the food category of the LIWC-22 dictionary. However, in the context of software development, ``cookie" typically refers to a type of data transmitted by a website to a user's web browser.

Additionally, it can be argued that, so far the assessment of LIWC's direct interpretability for psychological concepts and human factors has been mostly limited to general text. Therefore, to genuinely assess the tool's reliability for domain-specific text, it is imperative to replicate studies exploring the correlations between LIWC's output and psychological concepts using software engineering text such as, developers' mailing lists and code reviews. Such an approach could help mitigate the performance loss in tasks, such as personality detection, that involve understanding domain-specific communication nuances [S19]. 

Although our review primarily focused on concerns focused on LIWC’s use in software engineering, some studies have also noted broader ethical considerations when employing psychometric tools in this domain. For example, S19 acknowledges concerns regarding the use of personality detection techniques in SE research. While such concerns are not specific to LIWC itself, the use of automated psycholinguistic tools can raise important questions around privacy, consent, and the interpretation of psychological inferences. We acknowledge the importance of these issues and note that researchers should exercise care when applying such tools to sensitive or personal aspects of developer behavior.

\begin{mdframed}[backgroundcolor=lightgray!20,topline=false,leftline=false,rightline=false,bottomline=false] 
\textbf{RQ5 Summary}: We identify two types of concerns with the use of LIWC in SE: \textit{(a)} inherent limitations due to its lexicon-based nature (e.g., lack of negation handling, limited language support), and \textit{(b)} concerns with domain adoption  (e.g., absence and potential misrepresentation of SE-specific words).
\end{mdframed}

\section{Discussion}
\label{discussion}

Overall, our review indicates that LIWC has been successfully applied across a broad spectrum of SE tasks involving human factors, such as analyzing team dynamics, detecting emotions, and investigating various BSE concepts (e.g., leadership and personality). Studies have also demonstrated how some psycholinguistic frameworks from other domains that leverage LIWC can be adopted in SE studies, enriching SE research. However, several weaknesses were identified as well. Most notably, few papers rigorously evaluated the validity of LIWC’s outputs for their specific SE contexts; many appeared to assume that the tool’s general reliability would naturally extend to SE text. Furthermore, our findings show that no effort has been made to develop SE-specific dictionaries, which could address domain-related limitations and improve LIWC’s interpretability for SE-specific text. As a result, more validation efforts are needed to ensure that LIWC’s psychological categories can accurately assess SE text. In the remainder of this section, we explore some possibilities for improving and utilizing LIWC in novel ways for software engineering research.

\subsection{Software engineering specific dictionaries}
\label{se_dictionaries}
As detailed in Section \ref{inherent_limitations} certain concerns associated with LIWC are inherent in any lexicon-based approach while others are the result of adopting the tool in a new field. Among the latter concerns, the issue often brought up in the literature is the domain specificity of software engineering text and LIWC's limitations in capturing the text's context. The answer to this problem may lie in the creation of software-specific LIWC dictionaries which can prove to be a huge step forward toward a better understanding of developers and their unique linguistic patterns. While creating any LIWC dictionary demands a notable investment of time and effort, once created, it can present new opportunities for those studying developers' communications. Such dictionaries can be specific to particular settings in order to capture the language patterns in certain situations. For example, researchers interested in investigating the levels of stress and anxiety among developers could develop a custom dictionary that reflects the language use of developers in high-pressure situations. Furthermore, custom dictionaries could be defined with specific software engineering constructs, such as Team Management or Quality Assurance, in mind to more accurately capture the language patterns and psychological markers relevant to each activity. For example, dictionaries designed around team coordination could emphasize constructs like leadership, trust, or conflict resolution, enhancing the interpretability of LIWC's output in those contexts. Overall, LIWC possesses the potential to significantly improve our understanding of the text written by developers and therefore, addressing some of LIWC's limitations can prove to be invaluable. 

\subsection{Leveraging more LIWC categories for direct analysis of the text}
\label{more_categories}
Another way to improve the usability of LIWC is to take advantage of all its categories for direct analytical purposes. LIWC presents an expansive set of word categories. However, commonly, studies select specific categories, tailored to the needs of their particular analytical objectives, for close examination. As evident in Table \ref{tab:all_categories}, our findings point to the diversity of LIWC categories that have been utilized to serve different research purposes in software engineering studies. Yet, even though many software engineering studies have leveraged several of LIWC's capabilities, we still believe that the program holds untapped potential for conducting further text analysis through direct interpretation of its categories. Consider the category of ``assent'', known to correlate with agreement and passivity \cite{tausczik2010psychological}. Understanding assent can potentially help us in studying team conflicts, or analyzing discussions involving critical decision-making, such as tool adoption. Despite its potential relevance, this category has yet to be employed in order to assess agreement levels within developer conversations.

\subsection{LIWC and Large Language Models}
\label{liwc_llm}
With the rise of Large Language Models (LLM), researchers and practitioners are increasingly leveraging advanced machine learning-based methods to analyze text. However, unlike LIWC, LLMs operate as black-box models, making their decision-making processes difficult to interpret. While LLMs excel in generating human-like text and handling various NLP tasks, they often lack transparency when it comes to understanding how specific psychological or emotional dimensions in text are being analyzed. In contrast, LIWC offers a well-established, interpretable framework for extracting psycholinguistic insights, using predefined categories backed by years of validation in psychological research \cite{tausczik2010psychological}. This allows for clear, consistent analysis of psychological aspects in text, providing insights that are more directly interpretable and reliable for examining human factors in software engineering.

To overcome the interpretability challenges posed by black-box models, one promising approach involves integrating LIWC to enhance their ability to analyze text. As highlighted in our response to RQ1, LIWC categories have already been utilized as input in various machine learning models, often in combination with other features such as metadata from sources like Stack Overflow posts or GitHub repositories. By leveraging LIWC’s psychologically grounded insights, deep learning models could be enriched with valuable context that improves their ability to capture nuanced human emotions and cognitive states.

To overcome the interpretability challenges posed by black-box models, one promising approach involves integrating LIWC to enhance their ability to analyze text. As highlighted in our response to RQ1, LIWC categories have already been utilized as input in various machine learning models, often in combination with other features such as metadata from sources like Stack Overflow posts or GitHub repositories. LIWC features could even be incorporated into ensemble classifiers to improve performance across SE-related prediction tasks. By leveraging LIWC’s psychologically grounded insights, deep learning models could also be enriched with valuable context that improves their ability to capture nuanced human emotions and cognitive states. Given the challenges in interpretability that many deep learning models face \cite{doshi2017towards}, this hybrid approach offers a promising direction for future research and application.

Additionally, there is considerable room for analyzing LLM-generated content with the help of LIWC. Beyond evaluating the content of ChatGPT-generated responses, Kabir et al. analyzed the linguistic characteristics of such content generated in response to Stack Overflow questions and compared them to those written by humans [S43]. Their findings revealed that LLM-generated content tends to be more formal, analytic, and positive in language. These distinctions also suggest the potential for using LIWC to help differentiate between human- and LLM-authored text in software engineering artifacts. This kind of analysis can be extended to explore the linguistic features of different LLMs in various scenarios and conversations in software contexts. For instance, it would be insightful to determine what type of language is most effective in prompting an LLM when trying to perform programming tasks or address GitHub issues in a repository. Furthermore, it would be valuable to study how LLMs navigate conversations based on specific linguistic cues, whether they adjust their style in certain contexts, and how these adjustments impact the usefulness of their responses. 

Moreover, LLMs are susceptible to inherent biases in the text they generate, largely due to the vast, unfiltered datasets on which they are trained. LIWC can be instrumental in identifying such biases by analyzing language patterns in LLM-generated texts, particularly for sensitive categories related to gender, emotions, or ethnicity. For example, studies outside the software engineering domain have demonstrated gender differences in how LLMs replicate lexical, psychological, and social traits \cite{pervez2024inclusivity}. Applying these analyses to software engineering could reveal whether biases related to developers' identities emerge in LLM-generated responses in software engineering contexts e.g., addressing programming issues or performing tasks such as bug fixes or vulnerability detection. This could lead to important insights on how LLMs contribute to or mitigate biases in collaborative development environments.

\section{New Research Opportunities Leveraging LIWC}
LIWC offers exciting new possibilities that have not yet been extensively explored. Beyond the applications that we listed in Table \ref{tab:applications}, LIWC can be leveraged for several other tasks, such as:
(a) \textit{evaluating the quality and effectiveness of code comments}, by analyzing linguistic features to identify helpful and informative comments; (b) \textit{predicting project outcomes}, like delays or success, by analyzing linguistic patterns in communications; (c) \textit{identifying developer burnout}, by monitoring developer communication in project management tools and version control systems to detect signs of burnout or emotional stress among team members using categories such as \textit{Stress}, \textit{Social processes}, \textit{Assent}, etc.; (d) \textit{enhancing software documentation}, by identifying areas of improvements such as readability or ambiguity; and (e) \textit{improving product or app reviews}, by detecting linguistic patterns that reflect user satisfaction or indicate areas for improvement. Additionally, similar to the work by Janjua et al. [S41], LIWC can also be used to further our understanding of social engineering-based attacks in software projects by analyzing the communications between attackers and project members. While [S41] used LIWC categories as feature sets for models, a more direct linguistic analysis—focusing on the specific language patterns employed by attackers—can offer deeper insights into the behaviors and strategies behind social engineering attacks. This approach could help identify linguistic markers of manipulation, persuasion, and trust-building tactics in the context of these attacks.

Finally, beyond the immediate applications listed above, future studies can leverage LIWC in multiple ways to further strengthen behavioral software engineering research. For instance, developing custom SE-specific dictionaries around targeted psychological constructs or BSE concepts (see Section \ref{se_dictionaries}) could provide more precise insights into team dynamics or leadership behaviors. Researchers might also explore underutilized LIWC categories (as discussed in Section \ref{more_categories}) to capture new dimensions of developer communication that have not yet been investigated. Likewise, combining LIWC with deep learning or language models (Section \ref{liwc_llm}) may yield both improved classification performance and more interpretable results in tasks such as developer sentiment analysis or automated code review. By expanding these efforts, the community can continue pushing the boundaries of LIWC’s utility, creating a deeper, more contextualized understanding of human factors in software engineering.

\section{Threats to Validity}
\label{threats}

\textbf{Construct Validity}:
Papers retrieved from online libraries were manually filtered to select those pertinent to this study. While this method is susceptible to human error, we adopted a two-step process: one author conducted the primary selection, while another independently verified the final list. Although this adds a layer of validation, it does not fully replicate having two authors review every paper, and some risk of missing relevant studies remains. In terms of our search query, a potential threat to validity is the assumption that the naming conventions for LIWC are consistent across all relevant literature, which may overlook studies using alternate terminology. We mitigated this threat by explicitly checking for alternate naming conventions during our literature review but found none in the relevant studies across software engineering, psychology, and communication domains. In addition to assessing the title and abstract, we thoroughly review the entire paper before determining its inclusion in our study. This approach ensures the precision and relevance of our paper selection. To further mitigate bias, two authors were involved in the data extraction and coding phases of the study. We followed an iterative discussion and coding procedure until we reached substantial agreement for each category (above 0.6).

\textbf{Internal Validity}:
The internal operations of the online libraries we utilized in our study are undisclosed. Their methods of indexing studies and responding to search queries are not publicly available. Therefore it is possible, that this search process has potentially resulted in the omission of relevant studies. However, to mitigate this threat, we have collected relevant papers from six prominent and widely recognized online digital libraries, i.e., IEEE Xplore Digital Library, ACM Digital Library, Elsevier Science Direct, Springer Link Online Library, Wiley Online Library, and Scopus.

\textbf{External Validity}: 
In order to narrow down the survey to relevant papers, we implemented stringent inclusion and exclusion criteria. This led to our focus on papers written in English and accessible through one of our six chosen databases. However, this approach may have missed some papers, potentially affecting the study's generalizability. To mitigate this risk, we designed our search string to be as inclusive as possible, deliberately introducing more noise into our initial results. We later performed a meticulous manual analysis to eliminate irrelevant entries. This strategy aimed to ensure the inclusion of as many relevant papers as possible. Consequently, we invested significant time in the manual inspection process, resulting in the removal of over a thousand irrelevant papers.

\section{Conclusion}
The primary objective of this paper is to investigate the role of LIWC, a representative psycholinguistic tool widely recognized for its effectiveness in analyzing language and psychological patterns, in the software engineering research.  Through this study, we aim to identify LIWC's utility, outline its limitations and potential in SE research and practice. The selection of the 43 papers took place through a systematic process, using specific inclusion and exclusion criteria, as discussed Section \ref{methodology}. The findings from our five research questions, as summarized next, can serve as a potential resource for future work in the field, looking to utilize LIWC.

In addressing our first research question, we underlined the diverse purposes for which LIWC has been used within the software engineering domain. Our findings reveal an almost even split between studies that directly interpret LIWC's various dimensions and those that employ it in an indirect manner, often combined with other data or as input for machine learning models. We also highlight how, throughout the years, the community has demonstrated many unique methods of leveraging LIWC to address development-related issues and enhance the software engineering process. Furthermore, we present the types of data and data sources that were utilized in our selected studies, and leverage a taxonomy of software engineering activities adopted from previous research, to identify the activities that were most frequently studied using LIWC (e.g., \textit{Team Management} and \textit{Knowledge Management}). Additionally, by adopting Behavioral Software Engineering (BSE) concepts, we carefully annotated each paper based on the concepts it examined, offering a precise overview of the distribution of these studied concepts. This step allowed us to identify the concepts that were frequently studiedusing LIWC, as well as the ones that have received less attention or remained unexplored. Next, to better understand the potential and limitation of LIWC in analyzing domain-specific language used by software engineering practitioners, we investigated the evaluation methods used by researchers. Surprisingly, we discovered that a substantial portion of the papers neither directly evaluated LIWC's findings nor raised any concerns regarding its usage. Nonetheless, a noteworthy number of papers do subject LIWC's findings to different types of evaluations. We also note that ethical considerations, such as privacy and the potential for misinterpretation of psychometric inferences, are rarely addressed.

In summary, this study's findings demonstrate LIWC's unique ability to capture linguistic patterns that enhance our understanding of the human factors impacting software development and maintenance. While we believe that more of LIWC's capabilities remain to be explored, it is essential we understand the limitations of using an off-the-shelf textual analysis tool for SE-specific text. This study aims to shed light on the significance, applications, implications, and potential use of LIWC in  software engineering research. We hope that the insights from this work will empower future research to better recognize and leverage the possibilities offered by LIWC.

\appendix
\section{Selected Studies}
Below you can find a list of the studies used in this SLR:
\newline \newline
\indent [S1] Correa, D. and Sureka, A. (2014) ‘Chaff from the wheat: characterization and modeling of deleted questions on stack overflow’, in Proceedings of the 23rd international conference on World wide web. Seoul Korea: ACM, pp. 631–642. doi: 10.1145/2566486.2568036.
\newline

[S2] Licorish, S. A. and MacDonell, S. G. (2014a) ‘Personality profiles of global software developers’, in Proceedings of the 18th International Conference on Evaluation and Assessment in Software Engineering. London England United Kingdom: ACM, pp. 1–10. doi: 10.1145/2601248.2601265.
\newline

[S3] Calefato, F. et al. (2018) ‘Sentiment Polarity Detection for Software Development’, Empirical Software Engineering, 23(3), pp. 1352–1382. doi: 10.1007/s10664-017-9546-9.
\newline

[S4] Wang, Y. and Redmiles, D. (2016) ‘The Diffusion of Trust and Cooperation in Teams with Individuals’ Variations on Baseline Trust’, in Proceedings of the 19th ACM Conference on Computer-Supported Cooperative Work \& Social Computing. San Francisco California USA: ACM, pp. 303–318. doi: 10.1145/2818048.2820064.
\newline

[S5] Zolduoarrati, E. and Licorish, S. A. (2021) ‘On the value of encouraging gender tolerance and inclusiveness in software engineering communities’, Information and Software Technology, 139, p. 106667. doi: 10.1016/j.infsof.2021.106667.
\newline

[S6] Maier, T., DeFranco, J. and Mccomb, C. (2019) ‘An analysis of design process and performance in distributed data science teams’, Team Performance Management: An International Journal, 25(7/8), pp. 419–439. doi: 10.1108/TPM-03-2019-0024.
\newline

[S7] Rastogi, A. and Nagappan, N. (2016) ‘On the Personality Traits of GitHub Contributors’, in 2016 IEEE 27th International Symposium on Software Reliability Engineering (ISSRE). Ottawa, ON, Canada: IEEE, pp. 77–86. doi: 10.1109/ISSRE.2016.43.
\newline

[S8] Calefato, F. and Lanubile, F. (2022) ‘Using Personality Detection Tools for Software Engineering Research: How Far Can We Go?’, ACM Transactions on Software Engineering and Methodology, 31(3), pp. 1–48. doi: 10.1145/3491039.
\newline

[S9] Hergueux, J. and Kessler, S. (2022) ‘Follow the Leader: Technical and Inspirational Leadership in Open Source Software’, in CHI Conference on Human Factors in Computing Systems. New Orleans LA USA: ACM, pp. 1–15. doi: 10.1145/3491102.3517516.
\newline

[S10] Zolduoarrati, E., Licorish, S. A. and Stanger, N. (2022) ‘Impact of individualism and collectivism cultural profiles on the behaviour of software developers: A study of stack overflow’, Journal of Systems and Software, 192, p. 111427. doi: 10.1016/j.jss.2022.111427.
\newline

[S11] Wang, W. et al. (2020) ‘ArguLens: Anatomy of Community Opinions On Usability Issues Using Argumentation Models’, in Proceedings of the 2020 CHI Conference on Human Factors in Computing Systems. Honolulu HI USA: ACM, pp. 1–14. doi: 10.1145/3313831.3376218.
\newline

[S12] Mantyla, M. V. et al. (2017) ‘Bootstrapping a Lexicon for Emotional Arousal in Software Engineering’, in 2017 IEEE/ACM 14th International Conference on Mining Software Repositories (MSR). Buenos Aires: IEEE, pp. 198–202. doi: 10.1109/MSR.2017.47.
\newline

[S13] Gachechiladze, D. et al. (2017) ‘Anger and Its Direction in Collaborative Software Development’, in 2017 IEEE/ACM 39th International Conference on Software Engineering: New Ideas and Emerging Technologies Results Track (ICSE-NIER). Buenos Aires: IEEE, pp. 11–14. doi: 10.1109/ICSE-NIER.2017.18.
\newline

[S14] Raman, N. et al. (2020) ‘Stress and burnout in open source: toward finding, understanding, and mitigating unhealthy interactions’, in Proceedings of the ACM/IEEE 42nd International Conference on Software Engineering: New Ideas and Emerging Results. Seoul South Korea: ACM, pp. 57–60. doi: 10.1145/3377816.3381732.
\newline

[S15] Keertipati, S., Savarimuthu, B. T. R. and Licorish, S. A. (2016) ‘Approaches for prioritizing feature improvements extracted from app reviews’, in Proceedings of the 20th International Conference on Evaluation and Assessment in Software Engineering. Limerick Ireland: ACM, pp. 1–6. doi: 10.1145/2915970.2916003.
\newline

[S16] Júnior, M. C., Mendonça, M., Farias, M. A. D. F., Henrique, P., \& Corumba, D. (2012). ‘A neurolinguistic method for identifying OSS developers’ context-specific preferred representational systems’. In Proceedings of the Seventh International Conference on Software Engineering Advances. Lisbon, Portugal: IEEE, pp. 18–23.
\newline

[S17] Licorish, S. A. and MacDonell, S. G. (2018) ‘Exploring the links between software development task type, team attitudes and task completion performance: Insights from the Jazz repository’, Information and Software Technology, 97, pp. 10–25. doi: 10.1016/j.infsof.2017.12.005.
\newline

[S18] Brooks, I. and Swigger, K. (2012) ‘Using sentiment analysis to measure the effects of leaders in global software development’, in 2012 International Conference on Collaboration Technologies and Systems (CTS). Denver, CO, USA: IEEE, pp. 517–524. doi: 10.1109/CTS.2012.6261099.
\newline

[S19] Van Mil, F. C. J., Rastogi, A. and Zaidman, A. (2021) ‘Promises and Perils of Inferring Personality on GitHub’, in Proceedings of the 15th ACM / IEEE International Symposium on Empirical Software Engineering and Measurement (ESEM). Bari Italy: ACM, pp. 1–11. doi: 10.1145/3475716.3475775.
\newline

[S20] Bazelli, B., Hindle, A. and Stroulia, E. (2013) ‘On the Personality Traits of StackOverflow Users’, in 2013 IEEE International Conference on Software Maintenance. Eindhoven, Netherlands: IEEE, pp. 460–463. doi: 10.1109/ICSM.2013.72.
\newline

[S21] Hellman, J. et al. (2022) ‘Characterizing user behaviors in open-source software user forums: an empirical study’, in Proceedings of the 15th International Conference on Cooperative and Human Aspects of Software Engineering. Pittsburgh Pennsylvania: ACM, pp. 46–55. doi: 10.1145/3528579.3529178.
\newline

[S22] Licorish, S. A., Savarimuthu, B. T. R. and Keertipati, S. (2017) ‘Attributes that Predict which Features to Fix: Lessons for App Store Mining’, in Proceedings of the 21st International Conference on Evaluation and Assessment in Software Engineering. Karlskrona Sweden: ACM, pp. 108–117. doi: 10.1145/3084226.3084246.
\newline

[S23] Papoutsoglou, M., Kapitsaki, G. M. and Angelis, L. (2020) ‘Modeling the effect of the badges gamification mechanism on personality traits of Stack Overflow users’, Simulation Modelling Practice and Theory, 105, p. 102157. doi: 10.1016/j.simpat.2020.102157.
\newline

[S24] Han, Y. et al. (2023) ‘Cross-status communication and project outcomes in OSS development: A language style matching perspective’, Empirical Software Engineering, 28(3), p. 78. doi: 10.1007/s10664-023-10298-8.
\newline

[S25] Júnior, M. C. et al. (2021) ‘Industrial and OSS developers’ profiles: a family of experiments to evaluate a pioneering neuro-linguistic method for preferred representational systems automatic detection’, Journal of the Brazilian Computer Society, 27(1), p. 4. doi: 10.1186/s13173-021-00107-9.
\newline

[S26] Licorish, S. A. and MacDonell, S. G. (2014b) ‘Understanding the attitudes, knowledge sharing behaviors and task performance of core developers: A longitudinal study’, Information and Software Technology, 56(12), pp. 1578–1596. doi: 10.1016/j.infsof.2014.02.004.
\newline

[S27] Rigby, P. C. and Hassan, A. E. (2007) ‘What Can OSS Mailing Lists Tell Us? A Preliminary Psychometric Text Analysis of the Apache Developer Mailing List’, in Fourth International Workshop on Mining Software Repositories (MSR’07:ICSE Workshops 2007). Minneapolis, MN, USA: IEEE, pp. 23–23. doi: 10.1109/MSR.2007.35.
\newline

[S28] Almarimi, N. et al. (2023) ‘Improving the detection of community smells through socio‐technical and sentiment analysis’, Journal of Software: Evolution and Process, 35(6), p. e2505. doi: 10.1002/smr.2505.
\newline

[S29] Licorish, S. A. and MacDonell, S. G. (2013b) ‘Differences in Jazz project leaders’ competencies and behaviors: A preliminary empirical investigation’, in 2013 6th International Workshop on Cooperative and Human Aspects of Software Engineering (CHASE). San Francisco, CA, USA: IEEE, pp. 1–8. doi: 10.1109/CHASE.2013.6614725.
\newline

[S30] De Choudhury, M. and Counts, S. (2013) ‘Understanding affect in the workplace via social media’, in Proceedings of the 2013 conference on Computer supported cooperative work. San Antonio Texas USA: ACM, pp. 303–316. doi: 10.1145/2441776.2441812.
\newline

[S31] Licorish, S. A. and MacDonell, S. G. (2013c) ‘The true role of active communicators: an empirical study of Jazz core developers’, in Proceedings of the 17th International Conference on Evaluation and Assessment in Software Engineering. Porto de Galinhas Brazil: ACM, pp. 228–239. doi: 10.1145/2460999.2461034.
\newline

[S32] Licorish, S. A. and MacDonell, S. G. (2013a) ‘Adopting softer approaches in the study of repository data: a comparative analysis’, in Proceedings of the 17th International Conference on Evaluation and Assessment in Software Engineering. Porto de Galinhas Brazil: ACM, pp. 240–245. doi: 10.1145/2460999.2461035.
\newline

[S33] Licorish, S. A. and Macdonell, S. G. (2014) ‘Combining Text Mining and Visualization Techniques to Study Teams’ Behavioral Processes’, in 2014 IEEE 4th Workshop on Mining Unstructured Data. Victoria, BC, Canada: IEEE, pp. 16–20. doi: 10.1109/MUD.2014.10.
\newline

[S34] Licorish, S. A. and MacDonell, S. G. (2013d) ‘What Can Developers’ Messages Tell Us? A Psycholinguistic Analysis of Jazz Teams’ Attitudes and Behavior Patterns’, in 2013 22nd Australian Software Engineering Conference. Hawthorne, Victoria, Australia: IEEE, pp. 107–116. doi: 10.1109/ASWEC.2013.22.
\newline

[S35] Papoutoglou, M., Kapitsaki, G. M. and Mittas, N. (2018) ‘Linking Personality Traits and Interpersonal Skills to Gamification Awards’, in 2018 44th Euromicro Conference on Software Engineering and Advanced Applications (SEAA). Prague: IEEE, pp. 214–221. doi: 10.1109/SEAA.2018.00042.
\newline

[S36] Licorish, S. A. and MacDonell, S. G. (2012) ‘What affects team behavior? Preliminary linguistic analysis of communications in the Jazz repository’, in. (2012 5th International Workshop on Co-operative and Human Aspects of Software Engineering, CHASE 2012 - Proceedings), pp. 83–89. doi: 10.1109/CHASE.2012.6223029.
\newline

[S37] Licorish, S. A. and MacDonell, S. G. (2021) ‘Relating IS Developers’ Attitudes to Engagement’, in Proceedings of the 25th Australasian Conference on Information Systems (ACIS’14). 1–10.
\newline


[S38] Patel, P. et al. (2016) ‘Studying Expectation Violations in Socio-Technical Systems: a Case Study of the Mobile App Community’, in ECIS, p. ResearchPaper14.
\newline

[S39] Xia, X. et al. (2016) ‘It takes two to tango: Deleted stack overflow question prediction with text and meta features’, in 2016 IEEE 40th annual computer software and applications conference (COMPSAC). IEEE, pp. 73–82.
\newline

[40] Su’a, T. et al. (2017) ‘Quickreview: a novel data-driven mobile user interface for reporting problematic app features’, in Proceedings of the 22nd International Conference on Intelligent User Interfaces, pp. 517–522.
\newline

[S41] Janjua, F., et al. (2020). ‘Handling insider threat through supervised machine learning techniques’. In Procedia Computer Science, 177, pp. 64–71. doi: 10.1016/j.procs.2020.10.012
\newline

[S42] Salminen, J., et al. (2023). ‘How does an imaginary persona's attractiveness affect designers' perceptions and IT solutions? An experimental study on users' remote working needs’. In Information Technology \& People, 36(8), pp. 196–225. doi: 10.1108/ITP-09-2022-0729
\newline

[S43] Kabir, S., Udo-Imeh, D. N., Kou, B., \& Zhang, T. (2024). ‘Is Stack Overflow obsolete? An empirical study of the characteristics of ChatGPT answers to Stack Overflow questions’. In Proceedings of the CHI Conference on Human Factors in Computing Systems, pp. 1–17. doi: 10.1145/3613904.3642596
\newline

\section{PRISMA 2020 Items}
\label{app_b}

\renewcommand{\arraystretch}{1.2}
\begin{longtable}{|p{10cm}|p{3cm}|}
\caption{PRISMA 2020 Items with Reporting Locations} \label{tab:prisma_checklist} \\
\hline
\textbf{Checklist Item} & \textbf{Location Reported} \\
\hline
\endfirsthead

\hline
\textbf{Checklist Item} & \textbf{Location Reported} \\
\hline
\endhead

\hline
\endfoot

\hline
\endlastfoot

\textbf{Title:} Identify the report as a systematic review. & Title \\
\hline
\textbf{Abstract:} See the PRISMA 2020 for Abstracts checklist. & Abstract \\
\hline
\textbf{Introduction – Rationale:} Describe the rationale for the review in the context of existing knowledge. & Section \ref{sec1} \\
\hline
\textbf{Introduction – Objectives:} Provide an explicit statement of the objectives or questions the review addresses. & Section \ref{sec1} \\
\hline
\textbf{Methods – Eligibility criteria:} Specify inclusion and exclusion criteria and how studies were grouped. & Section \ref{manual_selection} \\
\hline
\textbf{Methods – Information sources:} Specify all sources searched and date of last search. & Sections \ref{query} \& \ref{manual_selection} \\
\hline
\textbf{Methods – Search strategy:} Present full search strategies for all sources. & Section \ref{methodology} \\
\hline
\textbf{Methods – Selection process:} Describe methods for deciding study inclusion and number of reviewers involved. & Section \ref{methodology} \\
\hline
\textbf{Methods – Data collection process:} Specify methods used to collect data. & Section \ref{data_extraction} \\
\hline
\textbf{Methods – Data items (Outcomes):} List and define all outcomes for which data were sought. & Section \ref{methodology} \\
\hline
\textbf{Methods – Data items (Other variables):} List and define all other variables for which data were sought. & N/A \\
\hline
\textbf{Methods – Study risk of bias assessment:} Methods used to assess risk of bias. & Section \ref{methodology} \\
\hline
\textbf{Methods – Effect measures:} Effect measures used for each outcome. & N/A \\
\hline
\textbf{Methods – Synthesis methods:} Describe methods used to synthesize results and explore heterogeneity. & Section \ref{methodology} \\
\hline
\textbf{Methods – Reporting bias assessment:} Describe methods used to assess risk of bias due to missing results. & N/A \\
\hline                   
\textbf{Results – Study selection:} Describe results of search and selection. & Section \ref{results} \\
\hline
\textbf{Results – Excluded studies:} Cite studies that may appear to meet inclusion criteria but were excluded. & Replication Package \\
\hline
\textbf{Results – Study characteristics:} Cite each included study and present characteristics. & Appendix A and Section \ref{results} \\
\hline
\textbf{Results – Results of individual studies:} Present summary statistics and effect estimates. & N/A \\
\hline
\textbf{Results – Results of syntheses:} Present results of all statistical syntheses, including heterogeneity. & N/A \\
\hline
\textbf{Discussion – Summary of evidence:} General interpretation of the results in the context of other evidence. & Section \ref{discussion} \\
\hline
\textbf{Discussion – Limitations of evidence:} Discuss limitations of included studies. & Section \ref{discussion} \\
\hline
\textbf{Discussion – Limitations of review process:} Discuss limitations of the review process. & \ref{threats} \\
\hline
\textbf{Discussion – Implications:} Discuss implications for practice, policy, and future research. & Section \ref{discussion} \\
\hline
\textbf{Other – Availability of data:} Report availability of data, code, and materials. & Section \ref{data_extraction} \\
\hline
\end{longtable}

 \bibliographystyle{elsarticle-num} 

\begin{thebibliography}{10}
\expandafter\ifx\csname url\endcsname\relax
  \def\url#1{\texttt{#1}}\fi
\expandafter\ifx\csname urlprefix\endcsname\relax\def\urlprefix{URL }\fi
\expandafter\ifx\csname href\endcsname\relax
  \def\href#1#2{#2} \def\path#1{#1}\fi

\bibitem{trinkenreich2022women}
B.~Trinkenreich, I.~Wiese, A.~Sarma, M.~Gerosa, I.~Steinmacher, Women’s participation in open source software: A survey of the literature, ACM Transactions on Software Engineering and Methodology (TOSEM) 31~(4) (2022) 1--37.

\bibitem{raman2020stress}
N.~Raman, M.~Cao, Y.~Tsvetkov, C.~K{\"a}stner, B.~Vasilescu, Stress and burnout in open source: Toward finding, understanding, and mitigating unhealthy interactions, in: Proceedings of the ACM/IEEE 42nd International Conference on Software Engineering: New Ideas and Emerging Results, 2020, pp. 57--60.

\bibitem{melnik2006comparative}
G.~Melnik, F.~Maurer, Comparative analysis of job satisfaction in agile and non-agile software development teams, in: International conference on extreme programming and agile processes in software engineering, Springer, 2006, pp. 32--42.

\bibitem{chatterjee2019exploratory}
P.~Chatterjee, K.~Damevski, L.~Pollock, V.~Augustine, N.~A. Kraft, Exploratory study of slack q\&a chats as a mining source for software engineering tools, in: 2019 IEEE/ACM 16th International Conference on Mining Software Repositories (MSR), IEEE, 2019, pp. 490--501.

\bibitem{SE_word_embeddings}
V.~Efstathiou, C.~Chatzilenas, D.~Spinellis, \href{https://doi.org/10.1145/3196398.3196448}{Word embeddings for the software engineering domain}, in: Proceedings of the 15th International Conference on Mining Software Repositories, MSR '18, Association for Computing Machinery, New York, NY, USA, 2018, p. 38–41.
\newblock \href {https://doi.org/10.1145/3196398.3196448} {\path{doi:10.1145/3196398.3196448}}.
\newline\urlprefix\url{https://doi.org/10.1145/3196398.3196448}

\bibitem{lyu2023detecting}
S.~Lyu, X.~Ren, Y.~Du, N.~Zhao, Detecting depression of chinese microblog users via text analysis: Combining linguistic inquiry word count (liwc) with culture and suicide related lexicons, Frontiers in psychiatry 14 (2023) 1121583.

\bibitem{bell2006variations}
C.~M. Bell, P.~M. McCarthy, D.~S. McNamara, Variations in language use across gender: Biological versus sociological theories, in: Proceedings of the Annual Meeting of the Cognitive Science Society, Vol.~28, 2006.

\bibitem{neysari2016monitoring}
M.~Neysari, G.~Bodenmann, M.~R. Mehl, K.~Bernecker, F.~W. Nussbeck, S.~Backes, M.~Zemp, M.~Martin, A.~B. Horn, Monitoring pronouns in conflicts, GeroPsych (2016).

\bibitem{simmons2005pronouns}
R.~A. Simmons, P.~C. Gordon, D.~L. Chambless, Pronouns in marital interaction: What do “you” and “i” say about marital health?, Psychological science 16~(12) (2005) 932--936.

\bibitem{ziemer2017using}
K.~S. Ziemer, G.~Korkmaz, Using text to predict psychological and physical health: A comparison of human raters and computerized text analysis, Computers in Human Behavior 76 (2017) 122--127.

\bibitem{gulliver2021assessing}
R.~Gulliver, K.~S. Fielding, W.~R. Louis, Assessing the mobilization potential of environmental advocacy communication, Journal of Environmental Psychology 74 (2021) 101563.

\bibitem{oc2023luxury}
Y.~Oc, K.~Plangger, S.~Sands, C.~L. Campbell, L.~Pitt, Luxury is what you say: Analyzing electronic word-of-mouth marketing of luxury products using artificial intelligence and machine learning, Psychology \& Marketing (2023).

\bibitem{lin2022opinion}
B.~Lin, N.~Cassee, A.~Serebrenik, G.~Bavota, N.~Novielli, M.~Lanza, Opinion mining for software development: a systematic literature review, ACM Transactions on Software Engineering and Methodology (TOSEM) 31~(3) (2022) 1--41.

\bibitem{edition2008psychology}
D.~W. Carroll, Psychology of language (2008).

\bibitem{harley2013psychology}
T.~A. Harley, The psychology of language: From data to theory, Psychology press, 2013.

\bibitem{levelt2013history}
W.~J. Levelt, A history of psycholinguistics: The pre-Chomskyan era, Oxford University Press, 2013.

\bibitem{traxler2011handbook}
M.~Traxler, M.~A. Gernsbacher, Handbook of psycholinguistics, Elsevier, 2011.

\bibitem{fletcher1995handbook}
P.~Fletcher, B.~MacWhinney, et~al., The handbook of child language, Blackwell Oxford, 1995.

\bibitem{miller2003cognitive}
G.~A. Miller, The cognitive revolution: a historical perspective, Trends in cognitive sciences 7~(3) (2003) 141--144.

\bibitem{harris1994chomskyan}
R.~A. Harris, The chomskyan revolution i: Syntax, semantics, and science, Perspectives on Science 2~(1) (1994) 38--75.

\bibitem{neuendorf2017content}
K.~A. Neuendorf, The content analysis guidebook, sage, 2017.

\bibitem{tausczik2010psychological}
Y.~R. Tausczik, J.~W. Pennebaker, The psychological meaning of words: Liwc and computerized text analysis methods, Journal of language and social psychology 29~(1) (2010) 24--54.

\bibitem{pennebaker2015development}
J.~W. Pennebaker, R.~L. Boyd, K.~Jordan, K.~Blackburn, The development and psychometric properties of liwc2015, Tech. rep. (2015).

\bibitem{boyd2022development}
R.~L. Boyd, A.~Ashokkumar, S.~Seraj, J.~W. Pennebaker, The development and psychometric properties of liwc-22, Austin, TX: University of Texas at Austin (2022) 1--47.

\bibitem{9402078}
P.~Chatterjee, K.~Damevski, L.~Pollock, Automatic extraction of opinion-based q\&a from online developer chats, in: 2021 IEEE/ACM 43rd International Conference on Software Engineering (ICSE), 2021, pp. 1260--1272.
\newblock \href {https://doi.org/10.1109/ICSE43902.2021.00115} {\path{doi:10.1109/ICSE43902.2021.00115}}.

\bibitem{sajadi2023interpersonal}
A.~Sajadi, K.~Damevski, P.~Chatterjee, Interpersonal trust in oss: Exploring dimensions of trust in github pull requests, in: 2023 IEEE/ACM 45th International Conference on Software Engineering: New Ideas and Emerging Results (ICSE-NIER), IEEE, 2023, pp. 19--24.

\bibitem{sajadi2023towards}
A.~Sajadi, K.~Damevski, P.~Chatterjee, Towards understanding emotions in informal developer interactions: A gitter chat study, in: Proceedings of the 31st ACM Joint European Software Engineering Conference and Symposium on the Foundations of Software Engineering, 2023, pp. 2097--2101.

\bibitem{imran2024uncovering}
M.~M. Imran, P.~Chatterjee, K.~Damevski, Uncovering the causes of emotions in software developer communication using zero-shot llms, in: Proceedings of the IEEE/ACM 46th International Conference on Software Engineering, ICSE '24, Association for Computing Machinery, New York, NY, USA, 2024.

\bibitem{felipe2023psychometric}
D.~A. Felipe, M.~Kalinowski, D.~Graziotin, J.~C. Natividade, Psychometric instruments in software engineering research on personality: Status quo after fifty years, Journal of Systems and Software 203 (2023) 111740.

\bibitem{yarkoni2010personality}
T.~Yarkoni, Personality in 100,000 words: A large-scale analysis of personality and word use among bloggers, Journal of research in personality 44~(3) (2010) 363--373.

\bibitem{10.1145/3611643.3613077}
R.~Ehsani, R.~Rezapour, P.~Chatterjee, \href{https://doi.org/10.1145/3611643.3613077}{Exploring moral principles exhibited in oss: A case study on github heated issues}, in: Proceedings of the 31st ACM Joint European Software Engineering Conference and Symposium on the Foundations of Software Engineering, ESEC/FSE 2023, Association for Computing Machinery, New York, NY, USA, 2023, p. 2092–2096.
\newblock \href {https://doi.org/10.1145/3611643.3613077} {\path{doi:10.1145/3611643.3613077}}.
\newline\urlprefix\url{https://doi.org/10.1145/3611643.3613077}

\bibitem{pardee1990motivation}
R.~L. Pardee, Motivation theories of maslow, herzberg, mcgregor \& mcclelland. a literature review of selected theories dealing with job satisfaction and motivation. (1990).

\bibitem{lenberg2015behavioral}
P.~Lenberg, R.~Feldt, L.~G. Wallgren, Behavioral software engineering: A definition and systematic literature review, Journal of Systems and software 107 (2015) 15--37.

\bibitem{sanchez2019taking}
M.~S{\'a}nchez-Gord{\'o}n, R.~Colomo-Palacios, Taking the emotional pulse of software engineering—a systematic literature review of empirical studies, Information and Software Technology 115 (2019) 23--43.

\bibitem{pennebaker2001linguistic}
J.~W. Pennebaker, M.~E. Francis, R.~J. Booth, Linguistic inquiry and word count: Liwc 2001, Mahway: Lawrence Erlbaum Associates 71~(2001) (2001) 2001.

\bibitem{kahn2007measuring}
J.~H. Kahn, R.~M. Tobin, A.~E. Massey, J.~A. Anderson, Measuring emotional expression with the linguistic inquiry and word count, The American journal of psychology 120~(2) (2007) 263--286.

\bibitem{lumontod2020seeing}
R.~Z. Lumontod~III, Seeing the invisible: Extracting signs of depression and suicidal ideation from college students’ writing using liwc a computerized text analysis, Int. J. Res. Stud. Educ 9 (2020) 31--44.

\bibitem{kangas2014can}
S.~E. Kangas, What can software tell us about political candidates?: A critical analysis of a computerized method for political discourse, Journal of Language and Politics 13~(1) (2014) 77--97.

\bibitem{tumasjan2010predicting}
A.~Tumasjan, T.~Sprenger, P.~Sandner, I.~Welpe, Predicting elections with twitter: What 140 characters reveal about political sentiment, in: Proceedings of the international AAAI conference on web and social media, Vol.~4, 2010, pp. 178--185.

\bibitem{apriyanto2020personality}
S.~Apriyanto, A.~Anum, Personality of politicians as the object of public assessment, in: Proceedings of the 2nd International Conference of Science and Technology for the Internet of Things, ICSTI 2019, September 3rd 2019, Yogyakarta, Indonesia, 2020.

\bibitem{stirman2001word}
S.~W. Stirman, J.~W. Pennebaker, Word use in the poetry of suicidal and nonsuicidal poets, Psychosomatic medicine 63~(4) (2001) 517--522.

\bibitem{pennebaker2008computerized}
J.~W. Pennebaker, C.~K. Chung, et~al., Computerized text analysis of al-qaeda transcripts, A content analysis reader 453465 (2008).

\bibitem{ashokkumar2021social}
A.~Ashokkumar, J.~W. Pennebaker, Social media conversations reveal large psychological shifts caused by covid-19’s onset across us cities, Science advances 7~(39) (2021) eabg7843.

\bibitem{kitchenham2009systematic}
B.~Kitchenham, O.~P. Brereton, D.~Budgen, M.~Turner, J.~Bailey, S.~Linkman, Systematic literature reviews in software engineering--a systematic literature review, Information and software technology 51~(1) (2009) 7--15.

\bibitem{page2021prisma}
M.~J. Page, J.~E. McKenzie, P.~M. Bossuyt, I.~Boutron, T.~C. Hoffmann, C.~D. Mulrow, L.~Shamseer, J.~M. Tetzlaff, E.~A. Akl, S.~E. Brennan, et~al., The prisma 2020 statement: an updated guideline for reporting systematic reviews, bmj 372 (2021).

\bibitem{batool2025ai}
A.~Batool, D.~Zowghi, M.~Bano, Ai governance: a systematic literature review, AI and Ethics (2025) 1--15.

\bibitem{ieeexplore}
IEEE, \href{https://ieeexplore.ieee.org/Xplore/home.jsp}{{IEEE Xplore Digital Library}} (2023).
\newline\urlprefix\url{https://ieeexplore.ieee.org/Xplore/home.jsp}

\bibitem{acmdigitallibrary}
ACM, \href{https://dl.acm.org/}{{ACM Digital Library}} (2023).
\newline\urlprefix\url{https://dl.acm.org/}

\bibitem{sciencedirect}
Elsevier, \href{https://www.sciencedirect.com/}{{ScienceDirect}} (2023).
\newline\urlprefix\url{https://www.sciencedirect.com/}

\bibitem{springerlink}
SpringerLink, \href{{https://link.springer.com/}}{Springerlink} (2023).
\newline\urlprefix\url{{https://link.springer.com/}}

\bibitem{wiley}
Wiley, \href{https://onlinelibrary.wiley.com/}{{Wiley Online Library}} (2023).
\newline\urlprefix\url{https://onlinelibrary.wiley.com/}

\bibitem{scopus}
Scopus, \href{https://www.scopus.com/home.uri}{Scopus} (2023).
\newline\urlprefix\url{https://www.scopus.com/home.uri}

\bibitem{halevi2017suitability}
G.~Halevi, H.~Moed, J.~Bar-Ilan, Suitability of google scholar as a source of scientific information and as a source of data for scientific evaluation—review of the literature, Journal of informetrics 11~(3) (2017) 823--834.

\bibitem{thomas2006general}
D.~R. Thomas, A general inductive approach for analyzing qualitative evaluation data, American journal of evaluation 27~(2) (2006) 237--246.

\bibitem{Landis77}
J.~R. Landis, G.~G. Koch, The measurement of observer agreement for categorical data, Biometrics 33~(1) (1977).

\bibitem{lenberg2015human}
P.~Lenberg, R.~Feldt, L.~G. Wallgren, Human factors related challenges in software engineering--an industrial perspective, in: 2015 ieee/acm 8th international workshop on cooperative and human aspects of software engineering, IEEE, 2015, pp. 43--49.

\bibitem{destefanis2016software}
G.~Destefanis, M.~Ortu, S.~Counsell, S.~Swift, M.~Marchesi, R.~Tonelli, Software development: do good manners matter?, PeerJ Computer Science 2 (2016) e73.

\bibitem{ortu2016emotional}
M.~Ortu, A.~Murgia, G.~Destefanis, P.~Tourani, R.~Tonelli, M.~Marchesi, B.~Adams, The emotional side of software developers in jira, in: Proceedings of the 13th international conference on mining software repositories, 2016, pp. 480--483.

\bibitem{souza2017sentiment}
R.~Souza, B.~Silva, Sentiment analysis of travis ci builds, in: 2017 IEEE/ACM 14th International Conference on Mining Software Repositories (MSR), IEEE, 2017, pp. 459--462.

\bibitem{pennebaker1999linguistic}
J.~W. Pennebaker, L.~A. King, Linguistic styles: language use as an individual difference., Journal of personality and social psychology 77~(6) (1999) 1296.

\bibitem{golbeck2011predicting}
J.~Golbeck, C.~Robles, K.~Turner, Predicting personality with social media, in: CHI'11 extended abstracts on human factors in computing systems, 2011, pp. 253--262.

\bibitem{goldberg1990alternative}
L.~R. Goldberg, An alternative" description of personality": the big-five factor structure., Journal of personality and social psychology 59~(6) (1990) 1216.

\bibitem{duijn2015quality}
M.~Duijn, A.~Kucera, A.~Bacchelli, Quality questions need quality code: Classifying code fragments on stack overflow, in: 2015 IEEE/ACM 12th Working Conference on Mining Software Repositories, IEEE, 2015, pp. 410--413.

\bibitem{druckman1988enhancing}
D.~Druckman, J.~A. Swets, et~al., Enhancing human performance: Issues, theories, and techniques, National Academies Press, 1988.

\bibitem{hofstede1984cultural}
G.~Hofstede, Cultural dimensions in management and planning, Asia Pacific journal of management 1 (1984) 81--99.

\bibitem{lavazza2010predicting}
L.~Lavazza, S.~Morasca, D.~Taibi, D.~Tosi, Predicting oss trustworthiness on the basis of elementary code assessment, in: Proceedings of the 2010 ACM-IEEE International Symposium on Empirical Software Engineering and Measurement, 2010, pp. 1--4.

\bibitem{calefato2017preliminary}
F.~Calefato, F.~Lanubile, N.~Novielli, A preliminary analysis on the effects of propensity to trust in distributed software development, in: 2017 IEEE 12th international conference on global software engineering (ICGSE), IEEE, 2017, pp. 56--60.

\bibitem{da2018arsenal}
G.~A.~M. da~Cruz, E.~H. Moriya-Huzita, V.~D. Feltrim, Arsenal-gsd: A framework for trust estimation in virtual teams based on sentiment analysis, Information and software technology 95 (2018) 46--61.

\bibitem{venigalla2021understanding}
A.~S.~M. Venigalla, S.~Chimalakonda, Understanding emotions of developer community towards software documentation, in: 2021 IEEE/ACM 43rd International Conference on Software Engineering: Software Engineering in Society (ICSE-SEIS), IEEE, 2021, pp. 87--91.

\bibitem{mohammad2013crowdsourcing}
S.~M. Mohammad, P.~D. Turney, Crowdsourcing a word--emotion association lexicon, Computational intelligence 29~(3) (2013) 436--465.

\bibitem{Likang2021}
L.~Yin, Z.~Chen, Q.~Xuan, V.~Filkov, \href{https://doi.org/10.1145/3468264.3468563}{Sustainability forecasting for apache incubator projects}, in: Proceedings of the 29th ACM Joint Meeting on European Software Engineering Conference and Symposium on the Foundations of Software Engineering, ESEC/FSE 2021, Association for Computing Machinery, New York, NY, USA, 2021, p. 1056–1067.
\newblock \href {https://doi.org/10.1145/3468264.3468563} {\path{doi:10.1145/3468264.3468563}}.
\newline\urlprefix\url{https://doi.org/10.1145/3468264.3468563}

\bibitem{4228660}
P.~C. Rigby, A.~E. Hassan, What can oss mailing lists tell us? a preliminary psychometric text analysis of the apache developer mailing list, in: Fourth International Workshop on Mining Software Repositories (MSR'07:ICSE Workshops 2007), 2007, pp. 23--23.
\newblock \href {https://doi.org/10.1109/MSR.2007.35} {\path{doi:10.1109/MSR.2007.35}}.

\bibitem{subash2022disco}
K.~M. Subash, L.~P. Kumar, S.~L. Vadlamani, P.~Chatterjee, O.~Baysal, Disco: A dataset of discord chat conversations for software engineering research, in: Proceedings of the 19th International Conference on Mining Software Repositories, 2022, pp. 227--231.

\bibitem{feldt2008towards}
R.~Feldt, R.~Torkar, L.~Angelis, M.~Samuelsson, Towards individualized software engineering: empirical studies should collect psychometrics, in: Proceedings of the 2008 international workshop on Cooperative and human aspects of software engineering, 2008, pp. 49--52.

\bibitem{katz1978social}
D.~Katz, R.~L. Kahn, The social psychology of organizations, Vol.~2, wiley New York, 1978.

\bibitem{isaksen2007assessing}
S.~G. Isaksen, G.~Ekvall, H.~Akkermans, G.~V. Wilson, J.~P. Gaulin, Assessing the Context for Change: A Technical Manual for the Situational Outlook Questionnaire, Enhancing Performance of Organizations, Leaders and Teams for Over 50 Years, Creative Problem Solving Group, 2007.

\bibitem{denison1996difference}
D.~R. Denison, What is the difference between organizational culture and organizational climate? a native's point of view on a decade of paradigm wars, Academy of management review 21~(3) (1996) 619--654.

\bibitem{seligman2000positive}
M.~E. Seligman, M.~Csikszentmihalyi, Positive psychology: An introduction., Vol.~55, American Psychological Association, 2000.

\bibitem{feisttheories}
J.~Feist, Theories of personality, brown \& benchmark, 1994, URL http://books. google. se/books.

\bibitem{WikipediaLeadershipPsychology}
{Wikipedia}, \href{https://en.wikipedia.org/wiki/Leadership_psychology}{Leadership psychology}, accessed in 2023.
\newline\urlprefix\url{https://en.wikipedia.org/wiki/Leadership_psychology}

\bibitem{forsyth2018group}
D.~R. Forsyth, Group dynamics, Cengage Learning, 2018.

\bibitem{feldman1984development}
D.~C. Feldman, The development and enforcement of group norms, Academy of management review 9~(1) (1984) 47--53.

\bibitem{messick1984nature}
S.~Messick, The nature of cognitive styles: Problems and promise in educational practice, Educational psychologist 19~(2) (1984) 59--74.

\bibitem{tett1993job}
R.~P. Tett, J.~P. Meyer, Job satisfaction, organizational commitment, turnover intention, and turnover: path analyses based on meta-analytic findings, Personnel psychology 46~(2) (1993) 259--293.

\bibitem{StudyComOrganizationalChange}
{study.com}, \href{https://study.com/learn/lesson/organizational-change-examples-theory-strategies.html}{Organizational change: Examples, theory \& strategies}, accessed in 2023 (2023).
\newline\urlprefix\url{https://study.com/learn/lesson/organizational-change-examples-theory-strategies.html}

\bibitem{vaughan2013social}
G.~M. Vaughan, M.~A. Hogg, Social psychology, Pearson Higher Education AU, 2013.

\bibitem{griesinger1973toward}
D.~W. Griesinger, J.~W. Livingston~Jr, Toward a model of interpersonal motivation in experimental games, Behavioral science 18~(3) (1973) 173--188.

\bibitem{APAStress}
{American Psychology Association}, accessed in 2023.
\newblock \href{https://dictionary.apa.org/stress}{[link]}.
\newline\urlprefix\url{https://dictionary.apa.org/stress}

\bibitem{tsai2010programmer}
M.-T. Tsai, N.-C. Cheng, Programmer perceptions of knowledge-sharing behavior under social cognitive theory, Expert Systems with Applications 37~(12) (2010) 8479--8485.

\bibitem{doshi2017towards}
F.~Doshi-Velez, B.~Kim, Towards a rigorous science of interpretable machine learning, arXiv preprint arXiv:1702.08608 (2017).

\bibitem{pervez2024inclusivity}
N.~Pervez, A.~J. Titus, Inclusivity in large language models: Personality traits and gender bias in scientific abstracts, arXiv preprint arXiv:2406.19497 (2024).

\end{thebibliography}

\end{document}